\documentclass[12pt]{article}
\usepackage[left=1.0in,right=1.0in,top=1.0in,bottom=1.0in]{geometry}
\usepackage{setspace}
\usepackage{fancyhdr}
\usepackage{lmodern}
\usepackage{microtype}
\usepackage[normalem]{ulem}
\usepackage[T1]{fontenc}

\usepackage{titlesec}
\usepackage[title]{appendix}
\usepackage{titling}
\pretitle{\begin{center}\large\bfseries}
\posttitle{\end{center}}
\preauthor{\begin{center}\normalsize}
\postauthor{\end{center}}
\predate{\begin{center}\normalsize}
\postdate{\end{center}}

\usepackage{amssymb, amsmath, amsfonts, mathtools}
\usepackage{dsfont}

\usepackage{array, booktabs, makecell}
\usepackage[flushleft]{threeparttable}
\usepackage{rotating, tabularx}

\usepackage{graphicx, subcaption}
\graphicspath{{figures/}}
\usepackage{pdflscape, tikz}
\usepackage{caption}
\usepackage{float}

\usepackage{enumitem}

\usepackage{xurl}
\usepackage{xcolor}
\definecolor{citationcolor}{RGB}{0, 127, 255}

\usepackage[backend=biber,
            style=authoryear,
            maxcitenames=3,
            mincitenames=1,
            maxbibnames=99,
            giveninits=true,
            uniquename=false,
            uniquelist=true,
            dashed=false,
            urldate=long,
            url=true,
            natbib=true]{biblatex}
\AtEveryBibitem{\clearfield{doi}\clearfield{eprint}\clearfield{issn}}

\DeclareCiteCommand{\cite}
  {\usebibmacro{prenote}}
  {\usebibmacro{citeindex}%
   \printtext[bibhyperref]{\color{citationcolor}\usebibmacro{cite}}}
  {\multicitedelim}
  {\usebibmacro{postnote}}
\DeclareCiteCommand{\parencite}[\mkbibparens]
  {\usebibmacro{prenote}}
  {\usebibmacro{citeindex}%
   \printtext[bibhyperref]{\color{citationcolor}\usebibmacro{cite}}}
  {\multicitedelim}
  {\usebibmacro{postnote}}

\newcolumntype{L}[1]{>{\raggedright\let\newline\\\arraybackslash\hspace{0pt}}m{#1}}
\newcolumntype{C}[1]{>{\centering\let\newline\\\arraybackslash\hspace{0pt}}m{#1}}
\newcolumntype{R}[1]{>{\raggedleft\let\newline\\\arraybackslash\hspace{0pt}}m{#1}}

\usepackage[hidelinks, breaklinks, colorlinks=true,
            linkcolor=citationcolor, citecolor=citationcolor,
            urlcolor=citationcolor]{hyperref}

\usepackage[nameinlink,capitalise]{cleveref}
\usepackage{multicol}
\crefname{section}{Appendix}{Appendices}
\Crefname{section}{Appendix}{Appendices}

  \title{Elitism in the Aisle: A Long-Run Surname Measure of\\ Legislative Elite Composition in Chile, 1834--2020\thanks{Replication data and code are available at \url{https://github.com/naimbro/elitism-in-the-aisle}. This work was funded by ANID---Millennium Science Initiative Program---code ICN17-002 and code 1523A0005 to N.B.\ and J.P.L., and by Fondecyt Iniciaci\'on 11251464. The funders had no role in study design, data collection and analysis, decision to publish, or preparation of the manuscript.}}
  \author{
  Naim Bro\thanks{Universidad Adolfo Ib\'a\~nez and Instituto Milenio Fundamentos de los Datos (IMFD). Email: \href{mailto:naim.bro.k@uai.cl}{naim.bro.k@uai.cl}.} \quad
  Juan Pablo Luna\thanks{McGill University; Pontificia Universidad Cat\'olica de Chile; and Instituto Milenio Fundamentos de los Datos (IMFD). Email: \href{mailto:juan.luna@mcgill.ca}{juan.luna@mcgill.ca}.}
  }

\date{\today}

\begin{document}

\maketitle
\thispagestyle{empty}

\begin{abstract}
\noindent \singlespacing
The link between descriptive and substantive representation is well established in the literature but is hard to trace historically, where class records are thin. We introduce a replicable \emph{enduring-elite surname measure}, pairing a contemporary socioeconomic criterion with historical elite registers, and apply it across the Chilean Congress, 1834--2020. Against a dynamic population reference built from 22.65 million birth registrations, the enduring-elite share of Congress falls from about half in the 1860s to about 12\% in the 2010s, with a sharp drop of 11 to 13 points around the 1925 constitutional reform. In 1910--1950, composition co-moves with the legislative agenda, net of party: common-surname legislators emphasize labor foremost, elite legislators a statecraft agenda of defense, foreign affairs, and administration. Across this window, who sits in Congress moves together with what Congress attends to.
\end{abstract}

\vspace{1em}

\newpage
\setcounter{page}{1}

\section{Introduction}\label{sec:intro}

The distinction between descriptive representation (who legislators are) and substantive representation (what they do) is foundational to the study of representation \parencite{Pitkin1967,Mansbridge1999}, and the link between the two is well documented: legislators from working-class backgrounds support greater state and social spending than their wealthier peers in Latin America \parencite{CarnesLupu2015} and the United Kingdom \parencite{OGrady2019}, social and group background shapes parliamentary debate more broadly \parencite{Fiva2021}, and these backgrounds shape preferences most strongly early in a career \parencite{CarnesLupu2023}. The difficulty is measuring social class. Knowing a legislator's social origin typically requires occupational records, surveys, or biographies, thin where they are most needed, over long runs of time and distant historical periods. The literature is thus overwhelmingly cross-sectional or short-span \parencite{Thompson2019,CarnesHansen2016,Bailer2022,BarnesSaxton2019,Murray2023,Gulzar2021}.

This article offers a method to close that gap: a replicable, long-range surname measure of legislative elite composition applied uniformly across nearly two centuries of the Chilean Congress, 1834--2020. We call it the \emph{enduring-elite surname measure}. The idea is simple: insofar as surnames carry ancestry, they carry social information, and a legislative body with many surnames that have persisted at the top of the hierarchy is more likely to be controlled by elites than one with few of them. The approach draws on an established literature that uses surnames to track social status and mobility across generations \parencite{Collado2008,Clark2015,Clark2014}, including in Chile and the Andean region \parencite{JaramilloEcheverri2021,BroMendoza2021} and for political alignments elsewhere \parencite{ByrneOMalley2013}, and on a parallel literature that reads names as socioeconomic signals of origin and status \parencite{FryerLevitt2004,AbramitzkyBoustanEriksson2014}. It is closest to, yet distinct from, work that identifies political dynasties from shared surnames \emph{within} a legislature \parencite{DalBoDalBoSnyder2009,CruzLabonneQuerubin2017,Smith2018}: that approach detects dynastic ties internally, whereas ours is an \emph{externally} anchored roster, tied to historical elite registers, that gauges the elite share of a whole chamber across two centuries rather than flagging individual dynasties. The measure is coarse: it separates an enduring elite from everyone else and cannot split the middle from the working class. But it is the same instrument in 1840 as in 2010, which makes it, to our knowledge, the longest-range estimate of legislative \emph{elite} composition of its kind. The dominant alternative codes legislators' occupations and class directly \parencite{CarnesLupu2015}; ours is blunter but cheaper and far longer-range.

The measure yields three payoffs, which organize the article. First, set against a dynamic population reference built from 22.65 million birth registrations, the elite share of Congress falls from a peak near 52\% in the 1860s to about 12\% in the 2010s, an over-representation of about 21-fold around 1900 that converges only slowly to about 7-fold by 2010. Second, this decline is not gradual: a sharp drop of 11 to 13 points coincides with the constitutional changes of 1925, the period historians of Chile label the breakthrough of the middle classes into Congress. Third, fitting a structural topic model to the corpus of congressional speeches, we show that between 1910 and 1950, the period of steepest change, elite and common legislators emphasize different topics, net of party: whereas the elite emphasizes defense, foreign affairs, and administration, the common group emphasizes labor, social welfare, and justice. In addition, a text-based ideal-point analysis on the same speeches places legislators on a single ideological axis, where the surname contrast points the same way but loses individual significance once chamber and tenure enter; we report that secondary stance analysis in the appendix.

\section{A replicable measure of legislative elite composition}\label{sec:measure}

The enduring-elite surname measure is built from two criteria: a surname must be elite \emph{now} and elite \emph{then}.

The first criterion identifies surnames at the top of the contemporary social hierarchy, by either of two routes. The first uses a surname-level socioeconomic index, a 0--100 score giving the average socioeconomic standing of the people who carry a given surname, estimated from the electoral roll by \textcite{BroMendoza2021}.\footnote{The index averages, across the people who carry a surname, the socioeconomic standing of their place of residence; \cref{app-sec:construction} gives the construction, following \textcite{BroMorav2024}.} Surnames of historically prominent landowning and commercial families (the Larra\'in, Edwards, Vial, and Subercaseaux lines) score near the top, whereas the most common Chilean surnames, such as Gonz\'alez and Mu\~noz, sit near the population average. A surname clears this route if its index is high (mean at least 65) and borne by enough people to be reliably estimated. The second route uses over-representation among contemporary business leaders: a surname clears it if it is (a) over-represented relative to its population frequency in a registry of 5,411 Chilean company directors and managers \parencite{CMF2023} and (b) historically attested in the elite registers below, among surnames borne by at least five legislators. This rule adds thirteen surnames to the socioeconomic list. These are recognizable enduring-elite lineages, among them the Balmaceda and Bulnes presidential families, the Portales line of the founding statesman of the Chilean republic, and Garc\'ia-Huidobro, which the socioeconomic route alone misses only because their surnames later spread across the population and diluted their average index; intersecting business over-representation with the historical registers recovers them while still excluding merely contemporary fortunes (\cref{app-subsec:bizroute}). The second criterion then applies to surnames passing either route.

The second criterion requires that the surname also appear in at least one historical elite source. These span the colonial period to the nineteenth century: the 1874 agricultural census, which records top-quartile landowners by valuation, together with registers of colonial nobility, prominent merchants, the 1882 enumeration of wealth holders, major shareholders, and mining elites. The full list of sources and the construction procedure are in \cref{app-sec:construction}.

The intersection of the two criteria is a set of \emph{298 enduring-elite surnames}. For each surname we also record a persistence-depth score, the number of historical registers in which it appears, from one up to ten.

To represent the population against which the elite is measured, we use the most frequent Chilean surnames, the \emph{common} surnames. At the person level, roughly half the population carries at least one common surname, which makes this group the right denominator for the elite share of Congress. Everything that is neither elite nor common falls into a residual \emph{other} category, a heterogeneous group of distinctive surnames broadly of middle and upper-middle origin, including faded historical families and post-1850 immigrant-origin families; it serves as the reference category in the regressions below.

\section{Validating the measure}\label{sec:validation}

The measure is built from surnames, so it must track elite status by independent criteria, not just by construction. We validate it against three kinds of external evidence: markers of elite social life from legislator biographies extracted from the National Congress Library, the structure of nineteenth-century family networks, and the known compositional difference between the two chambers.

\begin{figure}[t]
\centering
\includegraphics[width=0.9\linewidth]{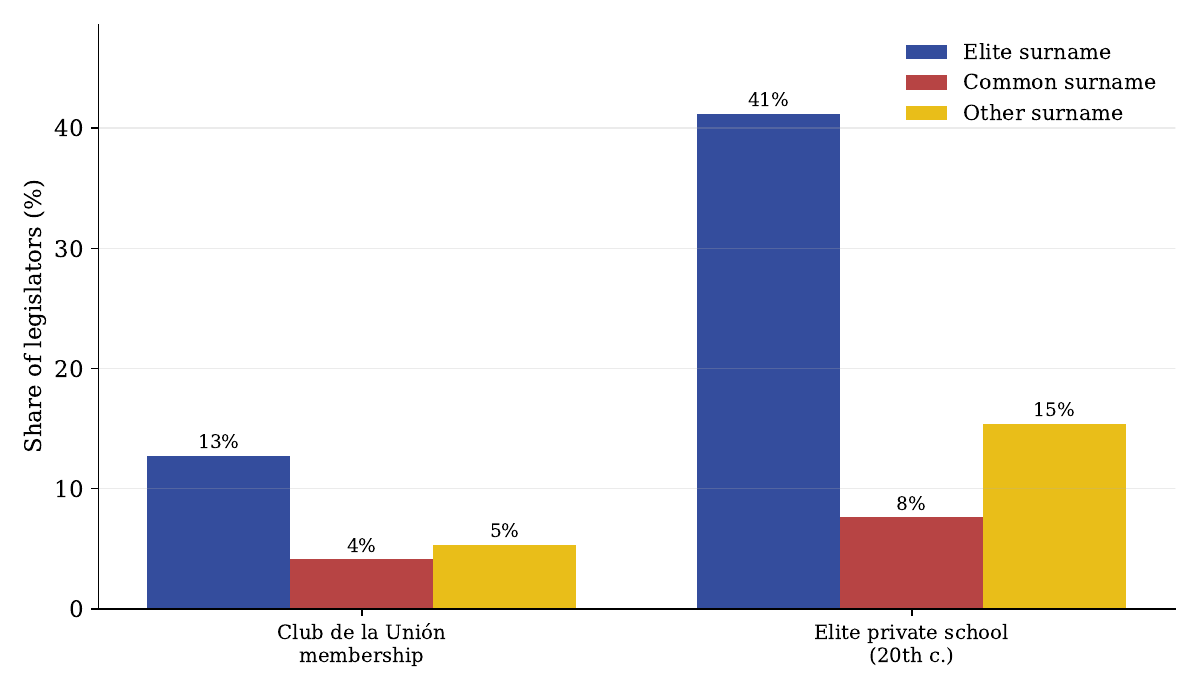}
\caption{Biographical markers of elite status by surname group. Bars give the share of legislators in each surname group (elite, common, and \emph{other}) who display each of two markers drawn from legislator biographies: membership in the elite Club de la Uni\'on and attendance at an elite twentieth-century private school. On both markers the common pole sits below the \emph{other} group and far below the elite pole. Source: legislator biographies (3,174 \emph{rese\~nas}, about 81\% of the roster) from the Library of Congress; group rates from replication output \texttt{should5\_common\_pole\_validation.csv}.}
\label{fig:validation}
\end{figure}

\cref{fig:validation} draws on two markers from legislator biographies (3,174 biographies, about 81\% of the roster, from the Library of Congress\footnote{The biographies were extracted from \url{https://www.bcn.cl/historiapolitica/} on 20 June 2026.}), broken out across the three surname groups. Membership in the elite Club de la Uni\'on runs at 12.7\% among elite-surname legislators, against 5.3\% in the \emph{other} group and 4.1\% among common-surname legislators;\footnote{The Club de la Uni\'on is a well-known elite space; see \textcite{Yeager1979}.} attendance at an elite twentieth-century private school runs at 41.2\%, against 15.4\% and 7.6\%. Biography coverage is balanced across surname groups (elite 80.5\%, common 80.0\%, other 82.0\%), so the gaps are not artifacts of differential coverage (\cref{app-tab:biocoverage}). On both markers the common pole sits below the \emph{other} group and far below the elite.

\begin{figure}[t]
\centering
\includegraphics[width=0.9\linewidth]{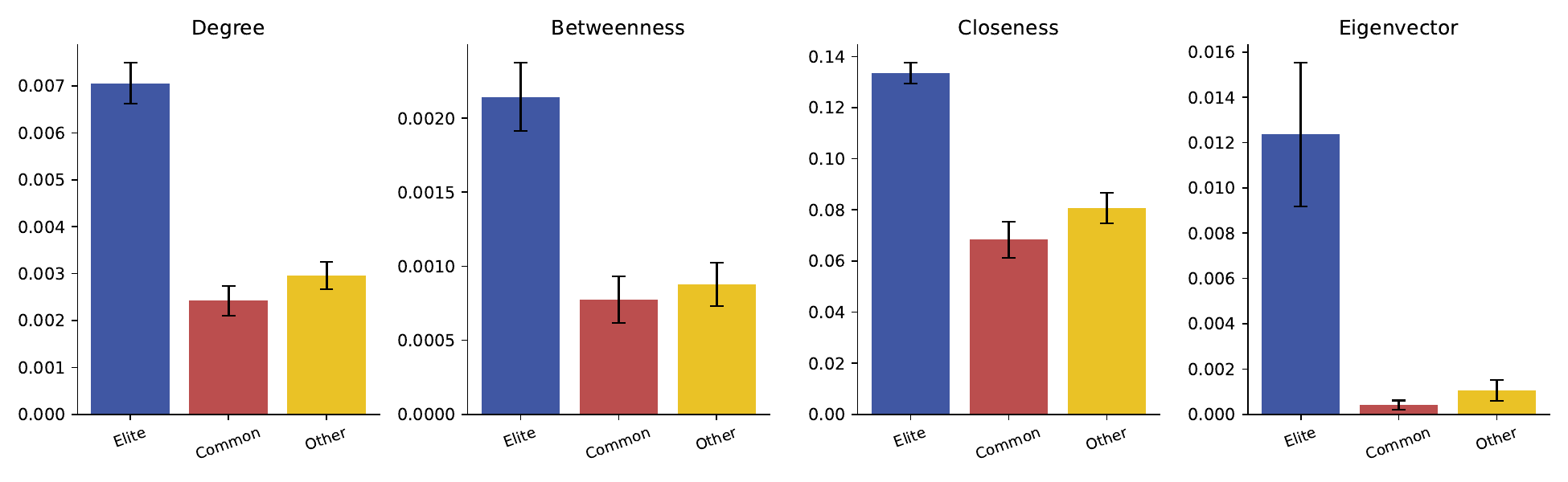}
\caption{Network centrality of legislators by surname group in the nineteenth-century genealogical network of the political class. Bars give mean centrality (with standard-error whiskers) for elite, common, and \emph{other} surname legislators on four measures: degree, betweenness, closeness, and eigenvector centrality. Source: genealogical kinship network of the Chilean political class \parencite{Bro2023}.}
\label{fig:centrality}
\end{figure}

\cref{fig:centrality} turns to the structural test, comparing the network centrality of elite-surname legislators with that of the others in the genealogical network of the nineteenth-century political class \parencite{Bro2023}. Elite-surname legislators are far more central: up to 11.8-fold on eigenvector centrality, and 2.4-fold by degree and 2.4-fold by betweenness. The families behind these surnames are not merely high-status today; they occupied the structural core of the political class when the historical criterion was being earned.

A further known-groups check works as expected: the Senate is more elite than the Chamber throughout. One caveat applies to the biographical markers: they are era-dependent, discriminating weakly in the nineteenth century, when nearly the whole chamber was aristocratic, and strongly in the twentieth, once it had diversified. This is a scope condition on the validation evidence, not on the measure itself. The full validation battery, with a note on how the biographies were extracted (a large language model for factual fields, class-coding by explicit researcher rules), is in \cref{app-sec:validation}.

\section{Nearly two centuries of legislative composition}\label{sec:composition}

\begin{figure}[t]
\centering
\includegraphics[width=0.9\linewidth]{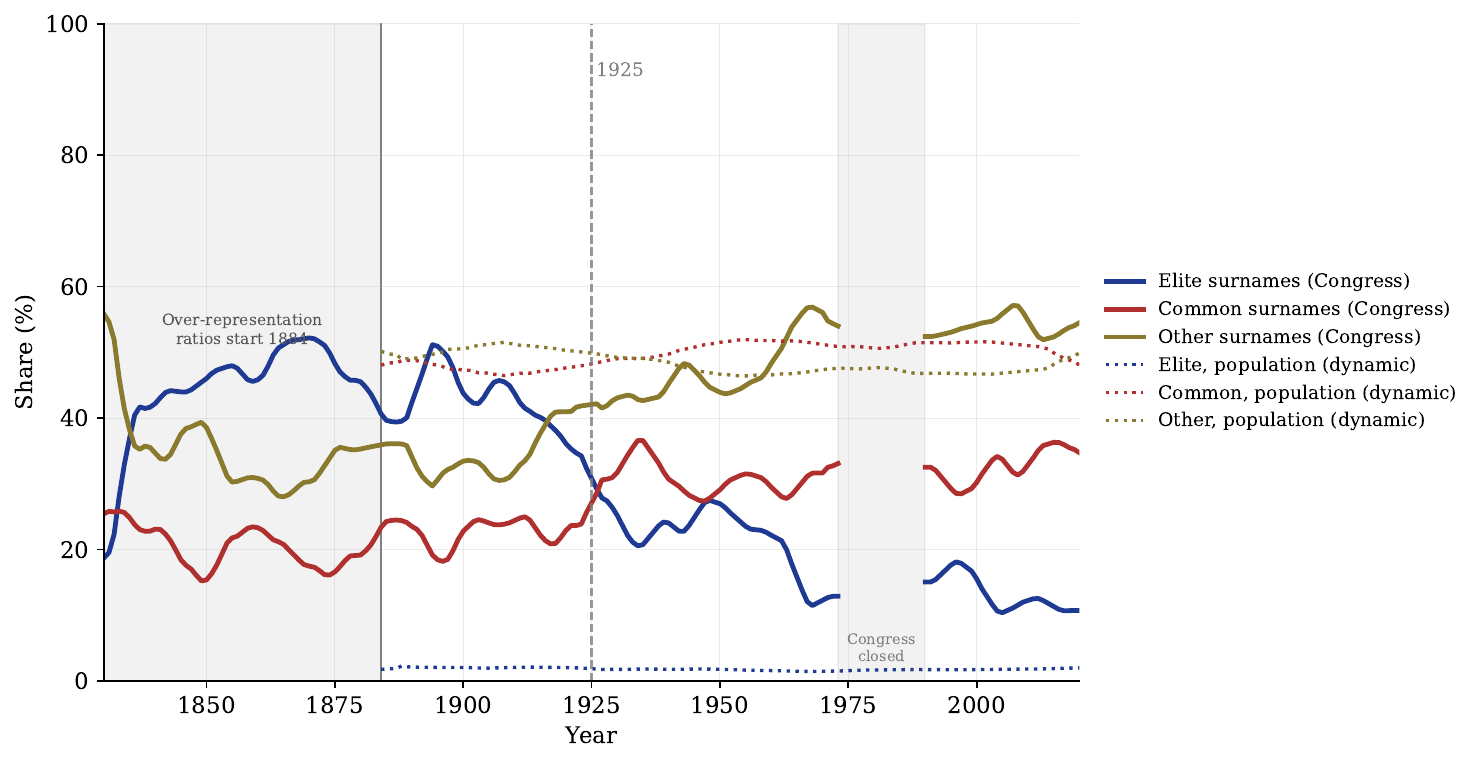}
\caption{Elite, common, and other surname shares of the Chilean Congress, 1834--2020, against a dynamic population reference.}
\label{fig:composition}
\end{figure}

\cref{fig:composition} applies the measure to every member of the Chilean Congress from 1834 to 2020 and sets the elite, \emph{other}, and common shares against a dynamic, person-level population reference. The Congress series runs the full span; the population reference begins in 1884, when civil registration was created in Chile.\footnote{The population reference is built from 22.65 million birth registrations of Chileans born from 1884 onward, obtained from the Servicio de Registro Civil e Identificaci\'on de Chile under Chile's Ley de Transparencia (transparency law). The person-level, surname-based construction follows \textcite{BroMorav2024}.} Because the reference is person-level, the benchmark moves with the actual surname stock of each decade rather than being fixed at a single year.

The trend is a long democratization of the legislature's surname stock. The elite share climbs from about 33\% in the 1830s to a peak near 52\% in the 1860s, then falls steadily to about 31\% in the 1920s and about 12\% in the 2010s; the common share rises symmetrically and crosses the elite line around 1925, as names such as Gonz\'alez, Rojas, and Mu\~noz take seats once held by Larra\'in, Vial, and Undurraga.

A third series, the residual \emph{other} surnames, rises materially across the same span, from about 31\% of Congress in the 1860s to about 55\% in the 2010s. Because that group gathers the lineages the measure cannot classify (faded historical families, immigrant-origin names, new commercial and industrial fortunes), part of the elite decline is the \emph{replacement} of an old elite by families the measure does not see, turnover as well as democratization.

Against a population in which elite surnames are a thin, stable slice (around 2\% throughout), this is dramatic over-representation: about 21-fold around 1900, falling to about 7-fold by 2010. Common-surname citizens run the other way, \emph{under}-represented throughout, from about 0.48-fold to only 0.64-fold by 2010. The two series converge toward less inequality of representation, but slowly, and never reach proportionality.

\section{A break in the 1920s: the decline of elite presence in Congress}\label{sec:break}

\begin{figure}[t]
\centering
\includegraphics[width=0.9\linewidth]{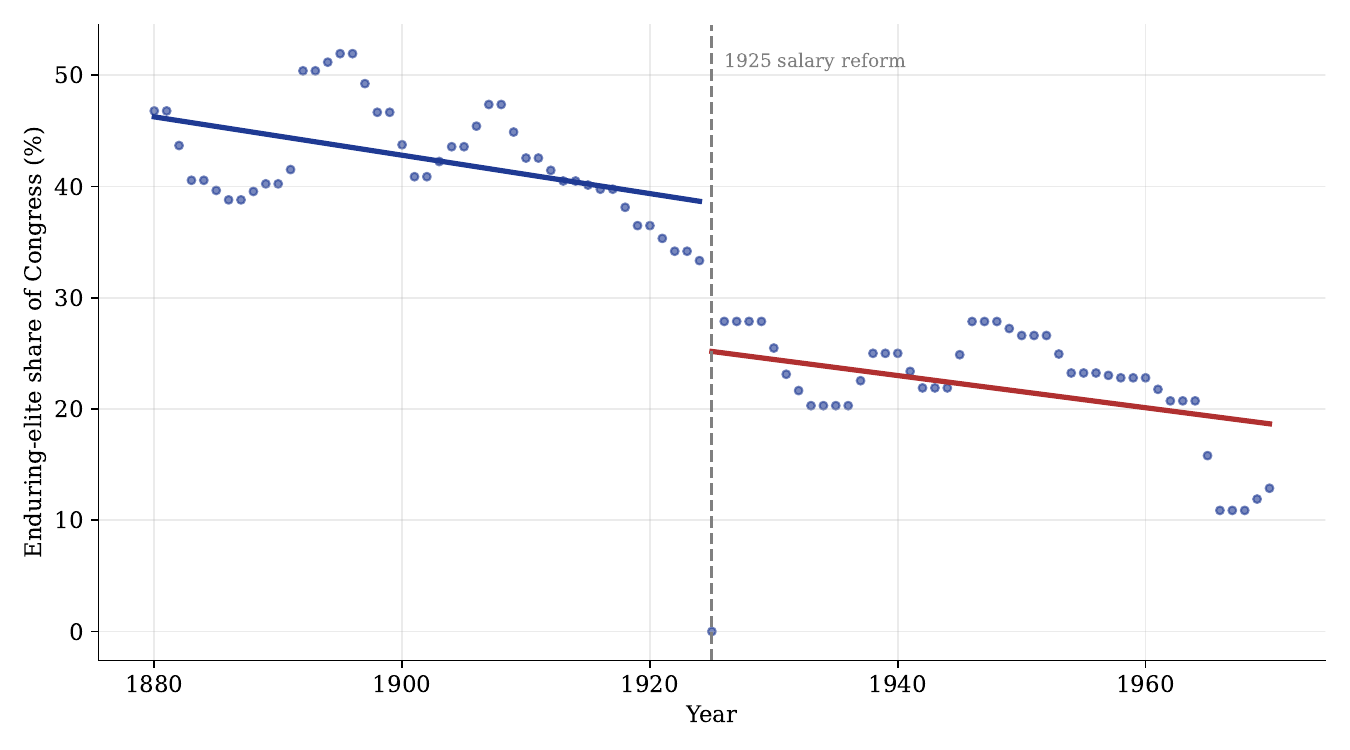}
\caption{Interrupted time series of the elite surname share of Congress around 1925. Points are the elite share of sitting legislators by year, 1880--1970; the fitted lines are a segmented (interrupted time-series) regression with the break set at 1925. Source: elite-surname coding of the full legislative roster (Section~\ref{sec:measure}).}
\label{fig:its1925}
\end{figure}

The compositional decline in \cref{fig:composition} is not smooth; it bends sharply, as an abrupt level shift, in the 1920s. \cref{fig:its1925} fits a segmented regression to the elite share of Congress over 1880--1970, with the break placed at 1925, the single best-fitting break year. Before the break, the elite share is already drifting down at $-0.17$ points per year. At 1925 the series drops by a level shift of $-13.3$ percentage points. We rest the inference on a placebo over break dates rather than on a $p$-value: across all candidate break years, the estimated level shift reaches its global maximum in magnitude at 1925, though only narrowly over the neighboring early-1920s years (\cref{app-fig:itsplacebo}), which on roughly ninety serially correlated annual proportions is more telling than the accompanying heteroskedasticity- and autocorrelation-consistent $p<0.001$. The elite-surname list and its construction are described in \cref{app-sec:construction}.

We read this break as one visible marker of a broader, well-documented process: the entrance of the middle classes into Congress in the 1920s and 1930s, long noted by historians of Chile \parencite{Correa2005,Salazar2015,Scully1992} and tied to the wider democratization of the period \parencite{CardosoFaletto1979}. The new Constitution of 1925 bundled many reforms at once (a constitutional rupture, a strengthened executive, the separation of church and state), among them a salary for members of Congress, which may have lowered the wealth barrier to a seat. Because these reforms were bundled, our design cannot say whether the step reflects the political incorporation of new sectors, the economic effect of paid legislative service, or both; we treat the break as coincident with that process rather than as evidence for any single mechanism.

The level shift holds near $-13$ points across windows from 1900--1955 to 1880--1970 (\cref{app-tab:itsrobust}). The 1925 maximum is a plateau rather than a knife-edge: the level shift is already near its maximum across the early 1920s (about $-13.0$ at 1922 and $-13.2$ at 1924) before peaking at 1925, so the design dates the elite decline to the early-1920s decade, not to 1925 precisely. It survives the surname coding rule too, holding negative and significant whether a legislator is coded elite by either surname, by both, or by the paternal surname alone (\cref{app-tab:codingrule}). The series regresses unweighted annual proportions, so the level break is a feature of the proportions themselves, not of any chamber-size weighting.

\section{Composition and agenda, 1910--1950}\label{sec:discourse}

A measure of who legislators are is worth only as much as its bearing on what they do. To test it, we turn to legislative speech in 1910--1950, the window of steepest compositional change in \cref{fig:composition}, recovered from the Library of Congress and converted to text by optical character recognition (116{,}938 speeches; the corpus and preprocessing are in \cref{app-sec:stm}). We fit a structural topic model \parencite{Roberts2014} with eighty topics to the speech corpus to recover what legislators talk about, and read the elite--common contrast in their agenda (the specification, preprocessing, and topic-count choice are in \cref{app-sec:stm}).

\begin{table}[!ht]
\centering
\begin{threeparttable}
\caption{Topic emphasis by surname group, aggregated to themes, legislative speech, 1910--1950}
\label{tab:themes}
\begin{tabular}{lccc}
\toprule
 & Elite & Common & Elite $-$ Common \\
Theme & emphasis & emphasis & contrast \\
\midrule
Elections \& parties      & $+0.99^{*}$  & $-0.15$      & $+1.13$ \\
Defense \& security       & $+0.29$      & $-0.14$      & $+0.43$ \\
State administration      & $+0.26$      & $-0.13$      & $+0.38$ \\
Foreign relations         & $+0.18^{**}$ & $-0.08$      & $+0.26$ \\
Education                 & $+0.03$      & $-0.04$      & $+0.07$ \\
Finance \& taxation       & $-0.15$      & $+0.15$      & $-0.29$ \\
Social welfare \& health  & $-0.06$      & $+0.25$      & $-0.31$ \\
Local government \& land  & $-0.48^{**}$ & $-0.18$      & $-0.30$ \\
Justice \& law            & $-0.03$      & $+0.38$      & $-0.41$ \\
Public works              & $-0.59^{**}$ & $-0.02$      & $-0.56$ \\
Labor \& industry         & $-1.05^{**}$ & $+0.72^{*}$  & $-1.77$ \\
\bottomrule
\end{tabular}
\begin{tablenotes}\small
\item \textit{Notes:} The 80 topics of the structural topic model \parencite{Roberts2014} are grouped into eleven themes; each cell is the summed party-, chamber-, and year-adjusted prevalence emphasis of the surname group relative to the \emph{other} reference (units of $10^{-2}$ of expected topic proportion), from document-level regressions with standard errors clustered by legislator. The final column is the elite-minus-common contrast (positive: elite emphasizes the theme more). Theme aggregation addresses the multiplicity of testing 80 topics; topic-level detail, labeling, and multiplicity are in \cref{app-sec:stm} and \cref{app-subsec:themes}. Stars mark each group's emphasis against the \emph{other} reference (first two columns), \emph{not} the contrast column the paper interprets; clustered $p$-values for the contrast, and the legislator-level slope tests that carry the agenda claim, are in \cref{app-subsec:themes}, \cref{app-tab:withinelite}, and \cref{app-tab:statecraft}. We read the ranking, not the magnitudes. Source: legislative speeches of the Chilean Congress, 1910--1950 (Library of Congress, OCR'd). $^{*}\,p<0.10$, $^{**}\,p<0.05$.
\end{tablenotes}
\end{threeparttable}
\end{table}

We group the fitted topics into eleven substantive themes and read the contrast at the theme level, which aggregates over topic granularity and so does not hinge on the topic count (\cref{tab:themes}; the topic-level detail and labeling procedure are in \cref{app-sec:stm}, the multiplicity treatment in \cref{app-subsec:themes}). Once party, chamber, and year are held fixed, the two groups bring different substance to the floor. Elite-surname legislators emphasize a statecraft cluster of elections and parties, national defense, state administration, and foreign relations. On the common side, labor and industry is the one clear positive emphasis, with social welfare and justice weakly so; the public-works and local-government themes sit at or below the \emph{other} reference and read as themes the elite \emph{de}-emphasizes rather than ones the common group actively presses. \cref{tab:themes} is a descriptive two-cluster ranking, not a set of single-theme tests: on the elite-minus-common contrast itself, only labor and industry, foreign relations, and elections and parties clear conventional thresholds under legislator clustering, and fewer still once the topic model's own estimation uncertainty enters (\cref{app-subsec:themes}, \cref{app-subsec:stminference}). We read the theme ranking, not the magnitudes, as the object of interest. The inferential weight of the agenda claim rests instead on the legislator-level slope tests below (\cref{app-tab:withinelite}, \cref{app-tab:statecraft}), which survive clustering and topic-model uncertainty.

The same speeches also place legislators on a single ideological axis, a text-based ideal-point model \parencite{Vafa2020} estimated from discourse rather than votes, and validated against the period's own party ordering (\emph{izquierda} $0.65 >$ \emph{centro} $0.44 >$ \emph{derecha} $0.13$). On that axis the surname contrast points the same way as the agenda (common-surname legislators lean more progressive, elite-surname legislators more conservative), but it is directional and party-robust rather than individually significant once chamber and tenure enter the preferred specification, so we report the ideal-point models, the axis validation, a legislator-level permutation test, and a contemporary NOMINATE convergent check in \cref{app-sec:tbip} as a secondary stance analysis rather than rest the paper's substantive claim on them.

The agenda contrast in \cref{tab:themes} is the payoff: where common surnames hold more seats, the floor turns toward labor and away from defense and statecraft, net of party. Who sits in Congress moves together with what Congress attends to.

One alternative deserves a direct test: the same window brought the FOCH labor federation, Recabarren's Socialist Workers' Party, and the Alessandri election of 1920, a broad opening of the agenda that might have moved elite legislators too. Fitting the same theme model within surname groups, labor-and-industry emphasis rises across the board from the 1910s to the 1940s, but significantly faster among common-surname legislators ($+1.5$ versus $+0.8$ points per decade, net of party and chamber; $p=0.002$ for the difference), so the elite--common gap widens rather than narrows (\cref{app-subsec:withinelite}). The widening is therefore inconsistent with a parallel secular shift in the agenda: as common surnames take seats, the gap grows rather than closes.

\section{Conclusion}\label{sec:conclusion}

This article builds a single instrument that reads the class composition of a legislature off the surnames of its members, and applies it to nearly two centuries of the Chilean Congress without changing the ruler from one decade to the next. The method is the contribution. Where the study of legislative class background has settled for cross-sections and short windows, the enduring-elite surname measure supplies a replicable, long-range alternative, putting the descriptive-to-substantive question on a footing that spans the period.

What it shows in Chile bears on that question. The decline of elite presence in Congress is real but slow, sharpest in the 1920s; and across the period of steepest change, the composition contrast lines up with an agenda contrast: where common surnames hold more seats, the floor turns more toward labor and social welfare, net of party.

The measure is coarse by design. It separates an enduring elite from everyone else, cannot tell a middle-class legislator from a working-class one, and its elite is, by construction, the one that endured to the present. Within those limits it is cheap, transparent, and reproducible, and its natural next use is comparative. Three conditions let it travel: surnames that transmit lineage, a long-run legislative roster, and at least one historical register that marks elite standing together with a contemporary status signal. Where those hold, the instrument can in principle be applied to another country's past, and one can ask whether the Chilean pattern, a long and uneven democratization of who sits in the chamber, is exception or rule.

\clearpage
{\centering\Large\bfseries Online Appendix\par}\vspace{1.5em}
\appendix
\setcounter{section}{0}\setcounter{table}{0}\setcounter{figure}{0}
\renewcommand{\thesection}{\Alph{section}}
\renewcommand{\thetable}{\Alph{section}\arabic{table}}
\renewcommand{\thefigure}{\Alph{section}\arabic{figure}}

\noindent This online appendix documents the construction of the enduring-elite surname measure, reports the full validation battery, and gives the underlying estimates behind the two discourse exhibits in the main text. It is supplementary, technical material: each section is self-contained and keyed to the relevant claim in the main text. All datasets and scripts named below are deposited in the public replication repository.

\vspace{0.5em}
\noindent The appendix is organized as follows. \Cref{app-sec:construction} sets out the two-criterion construction of the 298-surname measure and reports the convergent validity of the business-leader route. \Cref{app-sec:validation} gives the validation detail and the lesson we draw from the biographical-extraction step. \Cref{app-sec:stm} documents the structural topic model. \Cref{app-sec:tbip} reports the text-based ideal-point estimates. \Cref{app-sec:nominate} reports the contemporary convergent-validity check against NOMINATE. \Cref{app-sec:network} gives the genealogical network construction. \Cref{app-sec:gradient} reports the party class-gradient. \Cref{app-sec:robust} collects the robustness checks: the robustness of the 1925 break to window and break-year choice, a survivorship lower bound on the early elite share, the sensitivity of the composition results to the socioeconomic cutoff, and a theme-level reading of the topic contrast under multiplicity.

\section{Construction of the enduring-elite measure}\label{app-sec:construction}

The measure identifies surnames that are elite \emph{now} and elite \emph{then}. It is the intersection of two criteria applied to surnames; a surname enters the measure only if it clears both. The result is a list of \textbf{298 surnames}. The construction is implemented in scripts \texttt{01}, \texttt{02}, and \texttt{02b} of the replication repository; the final list is \texttt{enduring\_elite\_FINAL.csv}. This deposited 298-surname list, which incorporates the thirteen business-route additions and the manual reconciliation of source spellings, is canonical: it is the list used for every result in the paper, and it yields the headline 1925 level shift of $-13.32$ points. A purely deterministic, large-language-model-free rebuild of the list from the raw historical sources recovers most but not all of the deposited surnames and yields a slightly smaller break of about $-11.29$ points. That two-point gap is almost entirely the contribution of the thirteen business-leader-route surnames, not a spelling-reconciliation artifact: dropping those thirteen from the deposited list reproduces a break of $-11.13$ points, essentially the deterministic-rebuild figure (\cref{app-subsec:bizroutrobust}). The break is large and significant either way. We report all results on the deposited list.

\subsection{Criterion 1: elite now}

The first criterion flags surnames that sit at the top of the contemporary social hierarchy, by either of two routes. A surname clears the criterion if it satisfies \emph{either} route.

\textbf{Route 1a: surname socioeconomic index.} The first route uses the surname-level socioeconomic index of \textcite{BroMendoza2021}, a $0$--$100$ score giving the average socioeconomic standing of the people who carry a given surname, estimated from the 2020 electoral roll. The index is built in three steps: each individual on the roll is located to their place of residence; that locality is assigned a socioeconomic level from the 2012 Territorial Well-being Index, a census-based measure of the relative standing of each administrative unit down to the city block; and these individual values are averaged by surname. The construction follows \textcite{BroMorav2024}, who use the same residence-to-index linkage to measure the social standing of surname groups. Across the 84{,}693 surnames in the roll the index has a population mean of $49.9$. A surname clears this route if its index is at least $65$ and it is carried by at least $30$ bearers in the roll, so that the index is reliably estimated.

\textbf{Route 1b: overrepresentation among contemporary business leaders.} The second route uses a registry of $5{,}411$ Chilean company directors and managers, drawn from the public records of the Comisi\'on para el Mercado Financiero (CMF) and web-scraped on 13 January 2023 \parencite{CMF2023}. For each surname we count its appearances across both the paternal and the maternal slot, and we compute an overrepresentation ratio,
\[
\text{overrep}_s \;=\; \frac{\text{share of surname } s \text{ among business leaders}}{\text{share of surname } s \text{ in the population}},
\]
where the population denominator is aggregated from the full civil registry (about $45$ million surname tokens). A surname clears this route if it appears at least five times in the registry and its over-representation ratio is at least $8$ (its share among business leaders is at least eight times its population share). Surnames whose frequency among business leaders far exceeds their population frequency clear this route. Route 1b recovers surnames that have lost their socioeconomic-index signal because they have diluted across the population, but that remain concentrated among the contemporary business class.

\subsection{Criterion 2: elite then (required)}

The second criterion is required: a surname enters the measure only if it also appears in at least one nineteenth-century elite source. This anchors the surname to historical, not merely contemporary, elite standing, and it is what makes the estimand the \emph{enduring} elite. Because Criterion 2 is preserved for every surname, including those that enter through the business-leader route, the measure always captures families with documented nineteenth-century elite standing.

The Criterion-2 sources are a compilation of fourteen historical elite-family registers (colonial to nineteenth-century), parsed into surname lists:
\begin{enumerate}[topsep=2pt,itemsep=1pt]
  \item the \textbf{1874 agricultural census}, from which we take the surnames of top-quartile landowners ranked by land valuation; and
  \item thirteen further nineteenth-century registers documenting colonial nobility, prominent merchants of the early and mid-nineteenth century, the \textbf{1882 enumeration of major wealth holders}, the shareholders of early corporations, and the mining elites of the period.
\end{enumerate}
For each surname we record a \textbf{persistence-depth score}, the number of nineteenth-century sources in which the surname appears. The deepest-rooted surnames appear in many registers at once: Edwards in ten sources; Lyon, Matte, Ossa, and Cousi\~no in nine; Subercaseaux, Larra\'in, and Urmeneta in eight; Vicu\~na in seven. The persistence-depth score is a continuous companion to the binary measure; \cref{app-subsec:depthgate} uses it to re-run the composition trend and the 1925 break under a stricter gate that keeps only surnames attested in two or more registers.

\subsection{The business-leader route in operation}\label{app-subsec:bizroute}

The socioeconomic-index route alone misses a set of historically attested commercial-political families whose surnames have spread through the population, depressing their contemporary index even though the families remain prominent among business leaders. The business-leader route (Route 1b), intersected with the required Criterion 2, recovers \textbf{thirteen} such families: Garc\'ia-Huidobro, Gana, Pascual, Cox, Bulnes, Noguera, Balmaceda, Dom\'inguez, Grez, and Portales, together with Santa Cruz, James, and Salvador (the last three carrying no legislators in the 1910--1950 window). These are recognizable nineteenth-century political dynasties: Portales and Balmaceda alone name two of the defining political figures of the century. Because each cleared Criterion 2, adding them keeps the estimand the \emph{enduring} elite rather than the merely contemporary one. (Ruiz-Tagle, often discussed alongside these families, already entered the measure through the socioeconomic-index route and is not a business-route addition.)

\subsection{Convergent validity with business leaders}

\Cref{app-tab:bizvalid} reports the convergent validity of the measure against the business-leader registry. Of the 298 enduring-elite surnames, $52$ appear at least five times among the $5{,}411$ business leaders, and for those surnames the overrepresentation relative to population frequency is overwhelming and uniformly positive: not one of the in-measure surnames is under-represented among business leaders. Panel A lists the strongest in-measure cases; Panel B lists the thirteen surnames added through the business-leader route. The underlying counts and ratios are in \texttt{quality\_reports/business\_x\_historical.csv}.

\begin{table}[htbp]
\centering
\begin{threeparttable}
\caption{Convergent validity of the enduring-elite measure against contemporary business leaders}\label{app-tab:bizvalid}
\small
\begin{tabular}{lrr}
\toprule
Surname & Appearances among & Overrepresentation \\
        & business leaders  & ratio \\
\midrule
\multicolumn{3}{l}{\textit{Panel A: Strongest in-measure cases (already in the measure via the SES route)}} \\
\midrule
Matte        & 26 & 100.0$\times$ \\
Larra\'in    & 84 & 39.3$\times$ \\
Vial         & 54 & 36.0$\times$ \\
Err\'azuriz  & 39 & 51.9$\times$ \\
Claro        & 23 & 58.4$\times$ \\
Lyon         & 16 & 45.8$\times$ \\
Edwards      & 16 & 36.8$\times$ \\
Ossa         & 20 & 15.9$\times$ \\
Bezanilla    & 17 & 103.5$\times$ \\
Cruzat       & 25 & 21.7$\times$ \\
Undurraga    & 22 & 25.3$\times$ \\
Izquierdo    & 24 & 26.4$\times$ \\
\\[0.3em]
\multicolumn{3}{l}{\textit{Panel B: Surnames added through the business-leader route}} \\
\midrule
Garc\'ia-Huidobro & 13 & 139.5$\times$ \\
Gana              & 27 & 27.5$\times$ \\
Pascual           &  8 & 27.5$\times$ \\
Santa Cruz        & 10 & 25.5$\times$ \\
Cox               & 11 & 23.2$\times$ \\
Bulnes            & 13 & 22.0$\times$ \\
Noguera           &  8 & 16.4$\times$ \\
Balmaceda         &  7 & 11.0$\times$ \\
James             &  5 & 42.7$\times$ \\
Dom\'inguez       & 53 &  9.4$\times$ \\
Salvador          &  5 &  9.5$\times$ \\
Grez              & 10 &  8.9$\times$ \\
Portales          &  5 &  8.7$\times$ \\
\bottomrule
\end{tabular}
\begin{tablenotes}\small
\item \textit{Notes:} The table reports surnames that appear at least five times in a registry of $5{,}411$ Chilean company directors and managers (appearances counted across both the paternal and the maternal surname slot). The overrepresentation ratio is the surname's share among business leaders divided by its share in the population (the latter aggregated from the full civil registry, about $45$ million surname tokens). Of the 298 enduring-elite surnames, $52$ clear the five-appearance threshold; none is under-represented. Panel A reports a selection of the strongest in-measure cases, all of which already entered through the socioeconomic-index route. Panel B reports the thirteen surnames added through the business-leader route, all of which also clear the required historical criterion. Source: \texttt{quality\_reports/business\_x\_historical.csv}.
\end{tablenotes}
\end{threeparttable}
\end{table}

\subsection{The full 298-surname list}

\Cref{app-tab:fulllist} gives the complete list of the 298 enduring-elite surnames. The list and the construction scripts (\texttt{01}, \texttt{02}, \texttt{02b}) are in the public replication repository (\texttt{enduring\_elite\_FINAL.csv}). Surnames are rendered as recorded in the underlying electoral and historical sources (uppercase, accents stripped). The list spans several columns and runs across the page break below.

\begingroup
\renewcommand{\arraystretch}{1}
\captionof{table}{The full list of 298 enduring-elite surnames}\label{app-tab:fulllist}
\footnotesize
\begin{multicols}{4}
\raggedright
\begin{enumerate}[label=\arabic*.,leftmargin=2.4em,topsep=0pt,itemsep=0pt,parsep=0pt]
\item ABASCAL
\item ABASOLO
\item ACHURRA
\item ADAMS
\item ALCALDE
\item ALDUNATE
\item ALEMPARTE
\item ALESSANDRI
\item ALLEN
\item ALLIENDE
\item AMARAL
\item AMENABAR
\item AMENGUAL
\item AMOR
\item AMUNATEGUI
\item ANDONAEGUI
\item ANDUEZA
\item ANGUITA
\item ANINAT
\item ANTUNES
\item ANWANDTER
\item ARANGUA
\item ARIZTIA
\item ARMANET
\item ARMAS
\item ARRAU
\item ARRIARAN
\item ASPILLAGA
\item ASTABURUAGA
\item BAKER
\item BALBONTIN
\item BALMACEDA
\item BALTRA
\item BANADOS
\item BARANAO
\item BARAONA
\item BASTERRICA
\item BENAVENTE
\item BERNSTEIN
\item BESA
\item BESOAIN
\item BEZANILLA
\item BIANCHI
\item BOGGIANO
\item BONET
\item BRAND
\item BRIEBA
\item BRIEVA
\item BRISENO
\item BROWN
\item BROWNE
\item BRUNET
\item BUDGE
\item BULNES
\item BUNSTER
\item CALLEJAS
\item CALVO
\item CAMPBELL
\item CAMPINO
\item CAMPO
\item CANEPA
\item CANESSA
\item CANTOS
\item CAPO
\item CARDEMIL
\item CASANUEVA
\item CASTELLON
\item CLARO
\item COBO
\item CONDE
\item COOD
\item COSTA
\item COSTABAL
\item COSTAS
\item COUSINO
\item COX
\item CRUCHAGA
\item CRUZAT
\item CUADRADO
\item CUADROS
\item DAROCH
\item DARRIGRANDE
\item DELANO
\item DENEGRI
\item DEVES
\item DIEZ
\item DOMINGUEZ
\item DORADO
\item DRAGO
\item DUENAS
\item DUNCAN
\item EASTMAN
\item ECHAURREN
\item ECHAZARRETA
\item ECHENIQUE
\item EDWARDS
\item EGUIGUREN
\item ELGART
\item ELIZALDE
\item ELZO
\item ENRIQUEZ
\item ERRAZURIZ
\item ESCUTI
\item ESPOZ
\item ETCHEVERRY
\item EYZAGUIRRE
\item FAEZ
\item FERMANDOIS
\item FERRER
\item FERRIER
\item FONTAINE
\item FONTECILLA
\item FRANCE
\item GABLER
\item GALLEGO
\item GALLO
\item GANA
\item GANDARILLAS
\item GARCIA-HUIDOBRO
\item GARFIAS
\item GARMENDIA
\item GARNHAM
\item GARRETON
\item GARRIGA
\item GAZMURI
\item GEISSE
\item GIBBS
\item GREEN
\item GREZ
\item GROSSI
\item GUARACHI
\item GUEMES
\item GUMUCIO
\item GUNDELACH
\item GUNTHER
\item HALL
\item HEDERRA
\item HENDERSON
\item HERREROS
\item HILL
\item HITSCHFELD
\item HODGSON
\item HOTT
\item HUIDOBRO
\item IBIETA
\item IGLESIAS
\item IGUALT
\item INIGUEZ
\item INOSTROSA
\item IRARRAZAVAL
\item IZQUIERDO
\item JAMES
\item JEQUIER
\item JORDAN
\item JULIAN
\item JUSTINIANO
\item KENDALL
\item KENNEDY
\item KING
\item KLEIN
\item KRAUSE
\item LAMARCA
\item LAMAS
\item LARRAGUIBEL
\item LARRAIN
\item LARREA
\item LLONA
\item LOBO
\item LOIS
\item LYNCH
\item LYON
\item MACKENNA
\item MANTEROLA
\item MANZUR
\item MARFULL
\item MARTIN
\item MASCAYANO
\item MASSARDO
\item MATA
\item MATHIEU
\item MATTE
\item MAYOL
\item MENDIBURU
\item MILLER
\item MIQUEL
\item MOLLER
\item MONREAL
\item MONTT
\item MOORE
\item MORAL
\item MORANDE
\item MORRIS
\item MOURE
\item MULET
\item MULLER
\item NAREA
\item NAVAS
\item NEUMANN
\item NOGUERA
\item NUNO
\item O'BRIEN
\item OCHAGAVIA
\item OLAVARRIETA
\item OPASO
\item ORTUZAR
\item OSSA
\item OTERO
\item PALAZUELOS
\item PARGA
\item PARODI
\item PAROT
\item PASCUAL
\item PEIRANO
\item PERRY
\item PHILIPPI
\item PLATA
\item POLLONI
\item PONS
\item PORTALES
\item PRATS
\item PRICE
\item PUCCIO
\item PUELMA
\item QUESADA
\item RAPOSO
\item REDON
\item REEVES
\item RENARD
\item RENCORET
\item RICHARDS
\item RIESCO
\item RISOPATRON
\item RITCHIE
\item RONDANELLI
\item ROSENDE
\item ROSS
\item ROSSIER
\item ROVIRA
\item ROWE
\item RUIZ-TAGLE
\item RUSQUE
\item SALDES
\item SALVADOR
\item SANDERS
\item SANFUENTES
\item SANTA CRUZ
\item SANTAMARIA
\item SCHILLING
\item SCHNEIDER
\item SECO
\item SEGUI
\item SIERRALTA
\item SIMPSON
\item SMITH
\item SOFFIA
\item SOLARI
\item SOLER
\item SOLO
\item SORUCO
\item SOTA
\item SOTTA
\item SOUPER
\item SQUELLA
\item STARK
\item STUVEN
\item SUBERCASEAUX
\item SWETT
\item TAGLE
\item TALAVERA
\item TASSARA
\item TOCORNAL
\item TORNERO
\item TUNON
\item TUPPER
\item UNDURRAGA
\item UNZUETA
\item URMENETA
\item VALDIVIESO
\item VALLEDOR
\item VALLEJO
\item VELA
\item VELASCO
\item VIAL
\item VICUNA
\item VIEJO
\item VILDOSOLA
\item VILLOTA
\item WALKER
\item WARD
\item WATKINS
\item WERNER
\item WILSON
\item WOOD
\item YAVAR
\item ZALDIVAR
\item ZANARTU
\item ZEGERS
\item ZILLERUELO
\end{enumerate}
\end{multicols}
\par\vspace{0.4em}
\noindent\footnotesize\textit{Notes:} The complete set of 298 enduring-elite surnames, each clearing both the contemporary criterion (socioeconomic index or business-leader overrepresentation) and the required historical criterion. Surnames are listed alphabetically as recorded in the underlying electoral and historical sources. Source: \texttt{enduring\_elite\_FINAL.csv}; construction in replication scripts \texttt{01}, \texttt{02}, \texttt{02b}.
\endgroup
\normalsize

\section{Validation detail and the biographical-extraction lesson}\label{app-sec:validation}

This appendix gives the full validation battery summarized by the two validation figures in the main text, documents the source of the biographical markers, and draws an explicit methodological lesson from the extraction step.

\subsection{Criterion and network-centrality gaps}

The measure is built from surnames alone, so it must track elite status by independent criteria. On two biographical markers, elite-surname legislators are several times more likely than other legislators to display the trait:
\begin{itemize}[topsep=2pt,itemsep=1pt]
  \item \textbf{Membership in the elite Club de la Uni\'on}: $12.7\%$ against $4.9\%$.
  \item \textbf{Attendance at an elite twentieth-century private school}: $41.2\%$ against $12.5\%$.
\end{itemize}
On the genealogical kinship network of the nineteenth-century political class \parencite{Bro2023}, elite-surname legislators are far more central than other legislators: $11.8\times$ on eigenvector centrality, $2.4\times$ on degree, $2.4\times$ on betweenness, and $1.7\times$ on closeness. A further known-groups check works as expected: the Senate is more elite than the Chamber throughout the period, matching the conventional understanding of the two bodies.

\subsection{Source of the biographical markers}

The biographical markers come from legislator biographies (\emph{rese\~nas}) published by the Library of Congress. We recovered them through the Internet Archive's Wayback Machine, retrieving about $4{,}739$ archived pages and matching $3{,}174$ legislators, roughly $81\%$ of the full roster. From each biography we extracted a fixed set of fields (the father's surname, the legislator's university, club memberships, and schooling) using a batched large-language-model pass over the archived text. All biographical markers in this appendix, the elite-pole rates (Club de la Uni\'on $12.7\%$, elite school $41.2\%$) and the symmetric common-pole rates reported in the main text (Club de la Uni\'on $4.1\%$, elite school $7.6\%$) alike, are computed from this same scraped biographical corpus, which we are not licensed to redistribute. The corpus is therefore not deposited in the public replication repository, and these particular numbers are not regenerable from it; the repository instead documents the retrieval and extraction procedure so the markers can be reconstructed from the archived source by a user who re-scrapes it.

\subsection{Biography coverage by surname group}\label{app-subsec:biocoverage}

The biographical markers are observed only for legislators whose biography was recovered, so a natural worry is that coverage is correlated with surname status, inflating the fold-gaps in the main text. \Cref{app-tab:biocoverage} shows it is not. Coverage is balanced across surname groups: $80.5\%$ for elite-surname legislators, $80.0\%$ for common-surname legislators, and $82.0\%$ for the \emph{other} group, against an overall rate of $81.1\%$. Elite-surname legislators are, if anything, $0.6$ percentage points \emph{below} the average coverage rate, so coverage does not favor the elite. The two validation fold-gaps survive inverse-coverage reweighting to the full roster essentially unchanged: Club de la Uni\'on $2.6\times$ and elite school $3.3\times$. The Figure~1 fold-gaps are therefore not artifacts of differential biography coverage.

\begin{table}[htbp]
\centering
\begin{threeparttable}
\caption{Biography coverage by surname group}\label{app-tab:biocoverage}
\small
\begin{tabular}{lccc}
\toprule
Surname group & $N$ & Covered & Coverage (\%) \\
\midrule
Elite & 1034 & 832 & 80.5 \\
Common & 1070 & 856 & 80.0 \\
Other & 1813 & 1487 & 82.0 \\
\midrule
Overall & 3917 & 3175 & 81.1 \\
\bottomrule
\end{tabular}

\begin{tablenotes}\small
\item \textit{Notes:} Each row reports, for a surname group, the number of legislators in the roster ($N$), the number whose Library of Congress biography was recovered, and the resulting coverage rate. Coverage is balanced across groups (elite $80.5\%$, common $80.0\%$, other $82.0\%$; overall $81.1\%$), and elite-surname legislators are slightly below the average, so the validation fold-gaps are not artifacts of differential coverage. Reweighting the validation markers by the inverse of group coverage leaves the fold-gaps essentially unchanged (Club de la Uni\'on $2.6\times$, elite school $3.3\times$). Coverage is computed over unique legislator names; the covered count ($3{,}175$) exceeds the $3{,}174$ matched biographies reported in the main text because one biography matches two distinct roster name spellings. Source: match between the legislative roster and the recovered Library of Congress biographies; replication output \texttt{replication\_repo/output/bio\_coverage\_by\_group.csv}.
\end{tablenotes}
\end{threeparttable}
\end{table}

\subsection{Methodological lesson: extraction versus coding}\label{app-subsec:lesson}

We draw a deliberate line between extraction and coding. The large-language model is used \emph{only} for factual extraction: reading a field, such as the father's surname or the named university or club, off the biographical text. On these grounded fields the model is reliable, recovering the stated value at roughly $98$--$100\%$ accuracy against hand checks. The \emph{class coding}, the rules that translate an extracted field into an elite or non-elite category, is done in explicit, auditable researcher rules, not in model judgment.

This division was not abstract caution; it caught a concrete error. An early version that let the model classify schools coded the Instituto Nacional as an ``elite school.'' The Instituto Nacional is in fact a public, academically selective, meritocratic institution, not a marker of inherited elite standing. Correcting the coding by explicit rule did more than fix a mislabel: it surfaced the Instituto Nacional as a separable \emph{meritocratic public-school pathway} into the legislature, a route of achieved rather than inherited elite status. Had we delegated the class coding to the model, this distinction would have been silently collapsed.

One scope condition applies to the biographical markers, though not to the surname measure itself, which is built from the criteria alone. The markers are era-dependent. In the nineteenth century, when nearly the whole chamber was aristocratic, the markers discriminate weakly; in the twentieth century, once the legislature had diversified, they discriminate strongly. The validation evidence should be read in that light.

\section{Structural topic model}\label{app-sec:stm}

\subsection{Corpus and specification}

The structural topic model is estimated on legislative speeches from 1910--1950, recovered from the printed records of the Library of Congress and converted to text by optical character recognition with \texttt{pytesseract}. The unit of analysis is a single floor speech (an \emph{intervenci\'on}); speeches were tokenized and fragments shorter than about twenty tokens were dropped, leaving a document-term matrix of \textbf{116{,}938 documents and 20{,}685 terms}. The 1910--1950 window brackets the 1925 reform with roughly comparable context on either side and matches the span of machine-readable speech we were able to recover; OCR of century-old records is imperfect, so the corpus carries transcription noise, which the topic model absorbs as low-coherence topics that play no role in the contrast. We fit a structural topic model \parencite{Roberts2014} with \textbf{$K=80$ topics} and spectral initialization. The number of topics is not consequential for the substantive contrast, which we read at the theme level (\cref{app-subsec:stmtopics}, \cref{app-subsec:themes}). The prevalence formula is
\[
\texttt{\textasciitilde\ apellido + sector + camara + Year},
\]
where \texttt{apellido} is the surname group (elite, common, or the \emph{other} reference), \texttt{sector} the party bloc, \texttt{camara} the chamber, and \texttt{Year} the calendar year.

\subsection{Estimation of topic prevalence}

We separate the two stages of the model. The topic--word distributions are estimated \emph{once} from the corpus. Topic \emph{prevalence} by surname group is then obtained from \texttt{estimateEffect} with Global uncertainty, re-estimated under the 298-surname coding against the already-fitted model. We evaluate prevalence with party, chamber, and year held at their means, so the surname contrast is read net of party and chamber. This \texttt{estimateEffect} pass underlies the \emph{topic-level} illustration in \cref{app-tab:stmcoefs}; the \emph{theme-level} contrast of Table~1 in the main text is estimated separately, as the next subsection describes.

\subsection{Inference caveat}\label{app-subsec:stminference}

Two estimation paths underlie the speech results, with different inferential status. The \emph{topic-level} illustration in \cref{app-tab:stmcoefs} is read off \texttt{estimateEffect} with Global uncertainty; because those estimates are at the document level and a single legislator contributes many speeches that are not independent of one another, their standard errors understate sampling variability and the associated $p$-values overstate significance. We therefore report no stars on the topic-level table and read it as a descriptive illustration only. The \emph{theme-level} contrast of Table~1 is estimated separately: a document-level ordinary-least-squares regression of theme prevalence on surname group, party, chamber, and year, with standard errors \textbf{clustered by legislator} (\texttt{scripts/13\_topic\_theme\_effects.R}), which corrects for exactly that non-independence. The stars in Table~1 are therefore legislator-clustered and are the inferential basis for the main-text agenda contrast. The remaining caution at the theme level is multiplicity, which aggregating the eighty topics into eleven themes addresses (\cref{app-subsec:themes}); we accordingly read the theme \emph{ranking} rather than the individual magnitudes as the object of interest. The ideal-point analysis of \cref{app-sec:tbip} is a separate, secondary check on ideological \emph{stance}, not the inferential basis for the agenda contrast.

A further concern is that the theme proportions are themselves \emph{estimated} rather than observed: the legislator-clustered regression of Table~1 treats the structural topic model's fitted proportions as data, which corrects for non-independence but discards the topic model's own estimation uncertainty (a generated-regressand problem). We address it by the method of composition. We draw fifty samples of the theme proportions from the fitted model's posterior, re-estimate every theme contrast on each draw, and combine the draws by Rubin's rules (\texttt{scripts/28\_stm\_uncertainty\_methodofcomposition.R}), which propagates the topic-estimation uncertainty into the standard errors. The result is reassuring for the claim the paper actually makes and disciplining for the claim it does not. The two-cluster theme \emph{ranking} of Table~1 is unchanged: the statecraft themes (elections and parties, defense, state administration, foreign relations) stay on the elite side and the labor--local-infrastructure themes on the common side. Individual theme stars, by contrast, weaken once this extra uncertainty enters: the elections-and-parties and the common-side labor contrasts, for instance, no longer clear the $5\%$ threshold. This is exactly why the main text reads the two-cluster ranking, and not any single theme's magnitude, as the object of interest, and why no individual theme star is load-bearing. The same propagation applied to the within-group slope test of \cref{app-subsec:withinelite} leaves the common-minus-elite slope difference at $+0.84$ points per decade (combined $\mathrm{SE}=0.29$, $p=0.004$), with the topic-estimation component about $7\%$ of the total variance. The method of composition propagates uncertainty in the document--topic proportions ($\theta$) \emph{conditional on the fitted model}; it does not carry uncertainty in the topic--word distributions ($\beta$), the topic count $K=80$, or the spectral initialization. Robustness to those choices rests separately on the theme aggregation and the $K$-insensitivity argument of \cref{app-subsec:stmtopics}, which collapse topic granularity and so do not hinge on the particular fitted $\beta$ or $K$.

\subsection{Topic-level illustration, labeling, and selection}\label{app-subsec:stmtopics}

The main-text contrast is read at the theme level (Table~1). This subsection documents the topic-level detail beneath it: how the topics were labeled, which topics illustrate the contrast and why, and why the choice of $80$ topics does not drive the result.

\textit{Labeling.} The fitted model returns $80$ topics as ranked word lists. We labeled each by reading its highest-probability and FREX terms and assigning a short substantive gloss; the labels are descriptive summaries of those term lists, not modeling choices. The leading Spanish terms behind every displayed topic are given in \cref{app-tab:stmcoefs}, so each label can be checked against its words; the full $80$-topic term matrix is deposited in the replication repository.

\textit{Topic selection.} Several of the largest-coefficient topics are procedural or rhetorical (parliamentary speech formulae, committee mechanics, generic argumentation) rather than substantive policy. Because these carry no agenda content, \cref{app-tab:stmcoefs} displays the substantively interpretable policy topics on which the two groups differ most; the procedural and rhetorical topics are set aside, not because their gaps are small but because they are not interpretable as agendas. This is a curated illustration. The theme-level aggregation in Table~1 and \cref{app-subsec:themes} makes no such selection: it sums all $80$ topics into themes, so it is the selection-free version of the contrast and is the basis for the main-text claim.

\textit{Reading the table.} \Cref{app-tab:stmcoefs} reports each displayed topic's prevalence coefficient for the elite and common groups relative to the \emph{other} reference category, and the elite-minus-common contrast by which the table is organized. The elite-minus-common column is uniformly positive in the elite panel and negative in the common panel, consistent with the labels. Some elite-versus-\emph{other} coefficients are small or negative because the heterogeneous \emph{other} category itself contains high-status families; this is exactly why the main text reads the contrast as elite-versus-common rather than against \emph{other}.

\textit{Number of topics.} We fit $K=80$ topics. The substantive conclusion does not depend on this choice: aggregating to the eleven themes of Table~1 collapses the topic granularity, and the statecraft-versus-labor-and-local division is stable at that level. A larger or smaller $K$ would relabel and resplit topics but leave the thematic contrast intact; $K=80$ simply yields finer-grained topics to inspect.

\begin{table}[htbp]
\centering
\begin{threeparttable}
\caption{Topic-level illustration of the surname contrast (298-surname coding)}\label{app-tab:stmcoefs}
\small
\begin{tabular}{llrrr}
\toprule
Topic & Leading terms & Elite & Common & Elite $-$ \\
      &               & coef.       & coef.       & Common \\
\midrule
\multicolumn{5}{l}{\textit{Panel A: Policy topics the elite emphasizes more}} \\
\midrule
T29 & officers / army / navy / military        & $\phantom{-}0.0015$ & $-0.0013$ & $+0.0027$ \\
T1  & personnel / rank / staff / administrative& $\phantom{-}0.0006$ & $-0.0011$ & $+0.0018$ \\
T13 & minutes / convened / attended / adjourns & $-0.0005$           & $-0.0022$ & $+0.0017$ \\
T73 & electoral board / registration / lists   & $-0.0001$           & $-0.0012$ & $+0.0012$ \\
T42 & national / defense / war finance         & $-0.0005$           & $-0.0015$ & $+0.0010$ \\
\\[0.3em]
\multicolumn{5}{l}{\textit{Panel B: Policy topics the common group emphasizes more}} \\
\midrule
T63 & company / coal / mining / electricity    & $-0.0041$ & $-0.0013$ & $-0.0028$ \\
T55 & municipality / commune / department      & $-0.0015$ & $\phantom{-}0.0002$ & $-0.0017$ \\
T46 & court / judge / tribunal / justice       & $-0.0006$ & $\phantom{-}0.0009$ & $-0.0015$ \\
T59 & worker / organization / union / activity & $-0.0000$ & $\phantom{-}0.0013$ & $-0.0013$ \\
T35 & works / water / drinking water / sewerage& $-0.0003$ & $\phantom{-}0.0007$ & $-0.0010$ \\
\bottomrule
\end{tabular}
\begin{tablenotes}\small
\item \textit{Notes:} Topic-prevalence coefficients from \texttt{estimateEffect} with Global uncertainty, with party, chamber, and year held at their means, relative to the \emph{other} surname group. The final column is the elite-minus-common contrast that orders the table. These are the substantively interpretable policy topics on which the two groups differ most; procedural and rhetorical topics, some with larger raw gaps, are set aside (see the text above). This table is a curated illustration; the selection-free contrast is the theme-level Table~1 in the main text. Document-level estimates overstate significance because a legislator's speeches are not independent. Topic labels gloss the leading terms; the full term matrix is in the replication repository. Source: replication output \texttt{table1\_topic\_coefs\_298.csv}.
\end{tablenotes}
\end{threeparttable}
\end{table}

\section{Text-based ideal points}\label{app-sec:tbip}

\subsection{Method and sample}

We place legislators on a single discursive axis using the text-based ideal-point model of \textcite{Vafa2020}, which represents speeches as mixtures of latent topics and assigns each legislator an ideal point that bends those mixtures toward more left- or right-associated word choices. We estimate it with $50$ latent factors on a unigram vocabulary (Spanish stop words removed; terms in fewer than five speeches or more than $30\%$ of them dropped), optimizing the variational objective by stochastic gradient descent (Adam, learning rate $0.01$, batch size $1{,}024$, $12{,}000$ steps), with $\mathrm{Gamma}(0.3,0.3)$ priors on the nonnegative document and topic intensities and standard-normal priors on the ideological topics and the legislator ideal points. To estimate a stable ideal point we restrict the sample to legislators with at least $20$ speeches in the 1910--1950 window, since a legislator with only a handful of speeches carries too little text to be located reliably; this reduces the $597$ legislators with any speech to $281$ with enough text. The estimated ideal points range from $-0.84$ to $1.52$, with a mean of $0.25$ and a standard deviation of $0.83$. Higher values denote a more progressive discourse, an orientation we validate against the party ordering below.

\subsection{Axis validation}\label{app-subsec:axisvalid}

The progressive axis is not asserted but anchored in the period's own party spectrum. Mean ideal points by party bloc run
\[
\text{\emph{izquierda}}\;0.65 \;>\; \text{\emph{centro}}\;0.44 \;>\; \text{\emph{derecha}}\;0.13,
\]
so a higher score corresponds, by construction, to the more left-leaning end of the contemporaneous party spectrum. This ordering is what licenses reading the surname and status coefficients in \cref{app-tab:tbipfull} as movements toward or away from a more progressive discourse.

\subsection{Full ideal-point regressions}

\Cref{app-tab:tbipfull} presents five nested models of the legislator-level ideal point ($N=281$), and adds a sixth that replaces the categorical surname terms with the continuous surname socioeconomic index of \textcite{BroMendoza2021}. That index is the surname-level average of the present-day socioeconomic standing of the people who bear a surname: from the electoral roll, each individual is located to a census unit and assigned that unit's 2012 Territorial Well-being Index, and the values are averaged by surname (the construction follows \textcite{BroMorav2024}; see \cref{app-sec:construction}). Because it is a present-day measure applied to legislators of 1910--1950, it identifies a status effect only under the assumption that surname socioeconomic standing persists across the intervening generations, the same persistence the enduring-elite measure rests on, but here carried by a continuous contemporary score rather than by historical attestation. We therefore treat it as a robustness check on the categorical result, not as the headline. With that caveat: the categorical common-surname effect is positive throughout (common-surname legislators lean more progressive than the \emph{other} group), but specification-sensitive, and the elite-surname coefficient is negative and absorbed once party is controlled. The continuous index is the more precisely estimated of the two: a one-point increase in surname socioeconomic status is associated with a $0.0094$ lower ideal point (default SE $0.0035$, $p=0.007$), so higher status predicts a less progressive discourse, across specifications. It also survives clustering. \Cref{app-tab:tbipcluster} re-estimates the continuous specification with standard errors clustered by family: the coefficient is unchanged at $-0.0094$, the clustered standard error is $0.0039$ ($t=-2.39$, $p=0.018$ across 181 family clusters at $N=281$), and the effect remains significant at the 5\% level. Clustering at the family level is the demanding test here, because a legislator's surname status is shared with relatives in the chamber; the inference holds under it.

\begin{table}[htbp]
\centering
\begin{threeparttable}
\caption{Text-based ideal point on surname group and surname socioeconomic status, 1910--1950}\label{app-tab:tbipfull}
\small
\begin{tabular}{l cccccc}
\toprule
 & (1) & (2) & (3) & (4) & (5) & (6) \\
\midrule
Common surname (ref.\ other)   & $0.246^{**}$  & $0.216^{*}$   & $0.190$       & $0.177^{***}$ & $0.088$       &                \\
                               & $(0.120)$     & $(0.119)$     & $(0.118)$     & $(0.059)$     & $(0.055)$     &                \\
Elite surname (ref.\ other)    & $-0.200^{*}$  & $-0.093$      & $-0.136$      & $-0.009$      & $-0.030$      &                \\
                               & $(0.116)$     & $(0.119)$     & $(0.119)$     & $(0.059)$     & $(0.055)$     &                \\
Surname SES (continuous)       &               &               &               &               &               & $-0.009^{***}$ \\
                               &               &               &               &               &               & $(0.003)$      \\
Right (ref.\ center)           &               & $-0.253^{**}$ & $-0.264^{**}$ & $-0.036$      & $0.005$       & $-0.270^{**}$  \\
                               &               & $(0.121)$     & $(0.120)$     & $(0.060)$     & $(0.057)$     & $(0.118)$      \\
Left (ref.\ center)            &               & $0.188$       & $0.237$       & $-0.071$      & $-0.054$      & $0.203$        \\
                               &               & $(0.151)$     & $(0.153)$     & $(0.077)$     & $(0.072)$     & $(0.154)$      \\
Senate (ref.\ Chamber)         &               &               & $0.111$       & $0.028$       & $0.008$       & $0.114$        \\
                               &               &               & $(0.086)$     & $(0.043)$     & $(0.040)$     & $(0.085)$      \\
Years in Congress              &               &               & $0.009^{*}$   & $0.029^{***}$ & $0.024^{***}$ & $0.010^{**}$   \\
                               &               &               & $(0.005)$     & $(0.003)$     & $(0.003)$     & $(0.005)$      \\
Urban                          &               &               & $-0.082$      & $0.044$       & $0.001$       & $-0.089$       \\
                               &               &               & $(0.100)$     & $(0.050)$     & $(0.046)$     & $(0.100)$      \\
First year in Congress         &               &               &               & $0.054^{***}$ &               &                \\
                               &               &               &               & $(0.002)$     &               &                \\
\midrule
Decade fixed effects           & No            & No            & No            & No            & Yes           & No             \\
Observations                   & 281           & 281           & 281           & 281           & 281           & 280            \\
$R^2$                          & 0.044         & 0.083         & 0.110         & 0.781         & 0.821         & 0.116          \\
\bottomrule
\end{tabular}
\begin{tablenotes}\small
\item \textit{Notes:} OLS regressions of each legislator's text-based ideal point \parencite{Vafa2020} on surname group (reference: \emph{other}) or on the continuous surname socioeconomic index of \textcite{BroMendoza2021}. A higher ideal point denotes a more progressive discourse; the axis is validated against the party ordering (\emph{izquierda} $>$ \emph{centro} $>$ \emph{derecha}). The sample is the $281$ legislators with at least $20$ speeches ($280$ in column (6), where the continuous surname index is defined). Column (3) is the primary specification; column (4) adds an entry-cohort term, a post-treatment control that sharpens precision but over-controls; column (5) replaces it with decade fixed effects. Column (6) is the continuous-status specification. Standard errors in parentheses; clustering by legislator or by family leaves inference on the continuous index unchanged. * $p<0.10$, ** $p<0.05$, *** $p<0.01$.
\end{tablenotes}
\end{threeparttable}
\end{table}

\begin{table}[htbp]
\centering
\begin{threeparttable}
\caption{Continuous surname-status effect on the ideal point under family clustering}\label{app-tab:tbipcluster}
\small
\begin{tabular}{lcc}
\toprule
 & Default SE & Family-clustered SE \\
 & (1) & (2) \\
\midrule
Surname status (continuous) & -0.0094*** & -0.0094** \\
 & (0.0035) & (0.0039) \\
\midrule
$p$-value & 0.0071 & 0.0177 \\
Legislators & 281 & 281 \\
Clusters (families) & --- & 181 \\
\bottomrule
\end{tabular}

\begin{tablenotes}\small
\item \textit{Notes:} OLS regression of each legislator's text-based ideal point \parencite{Vafa2020} on the continuous surname socioeconomic index of \textcite{BroMendoza2021}, for the $281$ legislators with at least $20$ speeches in 1910--1950. Column (1) uses default (heteroskedasticity-robust) standard errors; column (2) clusters by family, where a family is the group of legislators sharing a paternal surname ($181$ clusters). A higher ideal point denotes a more progressive discourse. The point estimate is identical across columns; clustering inflates the standard error only modestly, and the effect remains significant at the five-percent level. * $p<0.10$, ** $p<0.05$, *** $p<0.01$. Source: replication output for the text-based ideal-point regressions.
\end{tablenotes}
\end{threeparttable}
\end{table}

\subsection{Permutation and clustered inference for the categorical contrast}\label{app-subsec:tbipperm}

The categorical common-versus-\emph{other} gap in the preferred specification (column 3 of \cref{app-tab:tbipfull}) is positive ($+0.19$) but not significant. \Cref{app-tab:tbipperm} confirms that null without leaning on the ordinary-least-squares standard error. The permutation test shuffles the surname labels across the $281$ legislators $10{,}000$ times, refits the preferred model on each draw, and reads the two-sided $p$-value off the resulting null distribution of the coefficient; it is therefore free of any parametric standard-error assumption. The permutation $p$-value for the common gap is $0.11$, essentially identical to the default OLS value, and family-clustering the same contrast by paternal surname leaves it at $0.13$. All three procedures agree: the categorical contrast is directional but not individually significant in the preferred specification. The paper's main-text substantive payoff rests on the legislative \emph{agenda} (the structural topic model of \cref{app-sec:stm}); these ideal-point results are reported as a secondary, corroborating check on ideological \emph{stance}, within which the continuous, family-clustered status result of \cref{app-tab:tbipcluster} is the more precisely estimated of the two.

\begin{table}[htbp]
\centering
\begin{threeparttable}
\caption{Permutation and clustered inference for the categorical surname contrast}\label{app-tab:tbipperm}
\small
\begin{tabular}{lcc}
\toprule
 & Common surname & Elite surname \\
 & (ref.\ other) & (ref.\ other) \\
\midrule
Coefficient (col 3 model) & $+0.190$ & $-0.136$ \\
Default OLS SE & (0.118) & (0.119) \\
Family-clustered SE & (0.126) & (0.135) \\
\midrule
$p$, default OLS & 0.108 & 0.253 \\
$p$, family-clustered & 0.132 & 0.312 \\
$p$, permutation & 0.109 & 0.242 \\
\midrule
Legislators & 281 & 281 \\
Clusters (families) & 181 & 181 \\
Permutations & 10,000 & 10,000 \\
\bottomrule
\end{tabular}

\begin{tablenotes}\small
\item \textit{Notes:} Inference for the categorical surname coefficients in the preferred specification (column 3 of \cref{app-tab:tbipfull}: ideal point on surname group, party, chamber, tenure, and an urban indicator), at the legislator level ($N=281$). The coefficient row repeats the column-3 point estimates. ``Default OLS'' uses heteroskedasticity-robust standard errors; ``family-clustered'' clusters by paternal surname ($181$ families). The permutation $p$-value is two-sided, computed from $10{,}000$ random reassignments of the surname labels across legislators, with an add-one correction; it requires no standard-error assumption. All three procedures agree that the common-surname gap is directional but not significant in the preferred specification. Source: replication output, \texttt{scripts/25\_tbip\_permutation.R}.
\end{tablenotes}
\end{threeparttable}
\end{table}

\subsection{Ideal points by surname group over time}

\begin{figure}[htbp]
\centering
\includegraphics[width=0.9\linewidth]{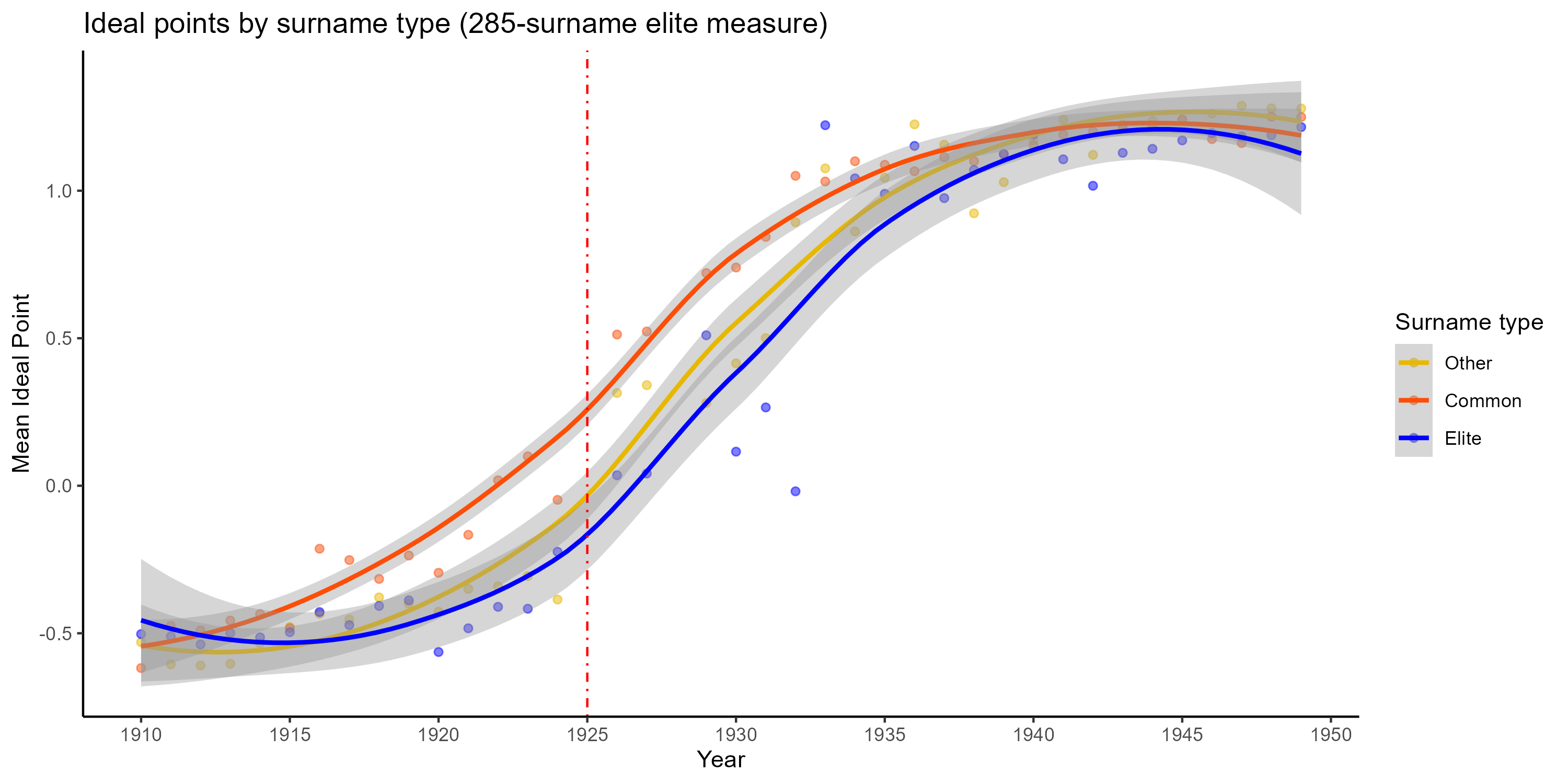}
\caption{Text-based ideal points by surname group over time, 1910--1950. Each point is a legislator's estimated ideal point \parencite{Vafa2020}, grouped by surname type (elite, common, \emph{other}), with higher values denoting a more progressive discourse on the period's own party-validated axis. Read the vertical separation between the surname groups as the discursive gap that persists within the period; the contrast tracks the regression estimates in \cref{app-tab:tbipfull}. Source: text-based ideal-point model fit to the 1910--1950 legislative-speech corpus (\texttt{puntos\_apellido\_revised.png}).}
\label{app-fig:tbiptime}
\end{figure}

\Cref{app-fig:tbiptime} plots estimated ideal points by surname group across the 1910--1950 window. It shows, period by period, that common-surname legislators sit at the more progressive end of the discursive axis and elite-surname legislators at the more conservative end, consistent with the regressions in \cref{app-tab:tbipfull}.

\section{Contemporary convergent validity: NOMINATE}\label{app-sec:nominate}

For recent congresses we check the measure against roll-call behavior rather than text, using first-dimension W-NOMINATE scores. Because W-NOMINATE recovers the polarity of each dimension arbitrarily from run to run, we re-orient each period so that the right is positive ($+$) before pooling.

\begin{figure}[htbp]
\centering
\includegraphics[width=0.9\linewidth]{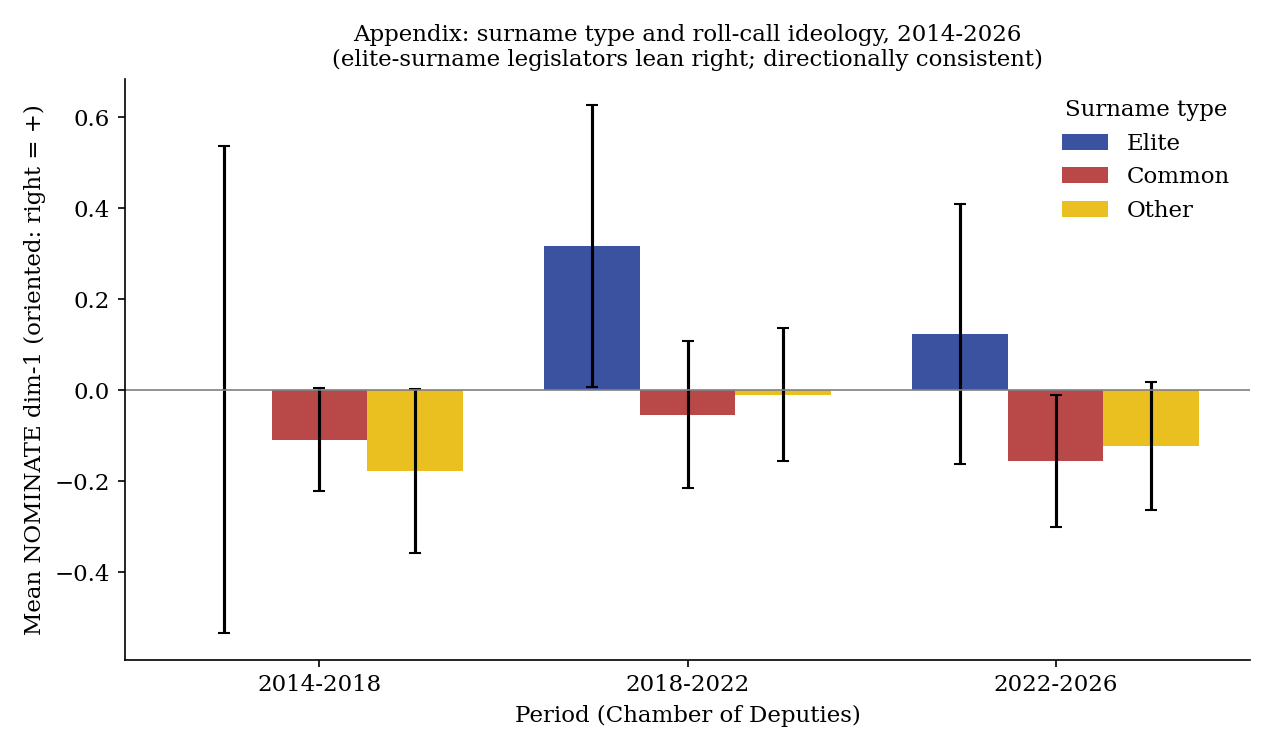}
\caption{Contemporary convergent validity against W-NOMINATE first-dimension scores. The figure plots the association between surname status and the re-oriented first NOMINATE dimension (right $=+$) across recent period-chambers. Read the estimates as the contemporary analogue of the 1910--1950 discourse gradient: the categorical elite-surname effect is directionally conservative but imprecise (pooled $+0.23$, $p=0.12$), reflecting the small number of elite-surname legislators ($\approx 75$) remaining in recent congresses, and the continuous-status effect is significant on its own ($p\approx0.001$) but absorbed by party ($p\approx0.42$). Source: W-NOMINATE scores for recent congresses (\texttt{figure\_appendix\_nominate.png}; replication script \texttt{15}).}
\label{app-fig:nominate}
\end{figure}

The categorical elite-surname effect on the first NOMINATE dimension is directionally consistent but underpowered. Pooled across recent period-chambers, an elite surname is associated with a $+0.23$ shift toward the right, but this is not significant ($p=0.12$). The imprecision is itself a consequence of the decline the paper documents: only about $75$ elite-surname legislators sit across the six recent period-chambers, so the contemporary elite cell is thin. A continuous surname socioeconomic measure is significant on its own ($p\approx0.001$): higher status predicts a more right-leaning roll-call record, but it \textbf{attenuates to approximately zero once party is controlled} ($p\approx0.42$). In other words, the contemporary status--ideology link runs \emph{through} party sorting, in contrast to the 1910--1950 period, when the status gradient in discourse held net of party. The relevant period-by-period estimates are plotted in \cref{app-fig:nominate}.

\section{Genealogical network centrality}\label{app-sec:network}

The structural validation reported in the main text (the network-centrality figure there) uses the nineteenth-century kinship network of the Chilean political class \parencite{Bro2023}, a genealogical graph of about $1{,}449$ nodes in which an edge records a documented family tie. On this network we compute four standard centrality measures: degree, betweenness, closeness, and eigenvector centrality, and compare group means across the elite, common, and \emph{other} surname groups.

Elite-surname legislators occupy the structural core of the network. Relative to other legislators, they are
\begin{itemize}[topsep=2pt,itemsep=1pt]
  \item $11.8\times$ more central by \textbf{eigenvector centrality},
  \item $2.4\times$ more central by \textbf{degree},
  \item $2.4\times$ more central by \textbf{betweenness}, and
  \item $1.7\times$ more central by \textbf{closeness}.
\end{itemize}
The very large eigenvector ratio is the telling one: elite-surname legislators are not only well-connected but connected to others who are themselves well-connected, which is what it means to sit at the core of a kinship network. The families behind the enduring-elite surnames thus occupied the structural center of the political class at exactly the time the measure's historical criterion was being earned.

\section{Party class-gradient, 1910--1950}\label{app-sec:gradient}

\Cref{app-tab:gradient} reports the surname composition of each party bloc across the 1910--1950 window under the 298-surname coding. The gradient is monotone and aligns with the conventional left--right ordering of the period's parties: right-of-center parties are elite-heavy and common-light, while left-of-center parties are common-heavy and elite-light. The Conservative and Liberal parties are roughly a fifth to a half elite by surname; the Radical, Socialist, and Communist parties are dominated by common surnames. The left parties contain \emph{few} (not zero) elite surnames: the elite share falls toward, but does not reach, zero on the left, consistent with a small number of elite-origin legislators crossing into reformist politics.

\begin{table}[htbp]
\centering
\begin{threeparttable}
\caption{Surname composition of party blocs, 1910--1950 (298-surname coding)}\label{app-tab:gradient}
\small
\begin{tabular}{lccc}
\toprule
Party bloc & Elite \% & Common \% & ($N$) \\
\midrule
Partido Conservador             & 48 & 17 & (497) \\
Partido Liberal                 & 43 & 22 & (429) \\
Partido Liberal Democr\'atico   & 34 & 26 & (155) \\
Partido Nacional (Monttvarista) & 41 & 28 & (68)  \\
Partido Democr\'atico           &  9 & 44 & (126) \\
Partido Radical                 & 15 & 33 & (571) \\
Partido Socialista              & 15 & 43 & (68)  \\
Partido Comunista               &  2 & 44 & (55)  \\
\bottomrule
\end{tabular}
\begin{tablenotes}\small
\item \textit{Notes:} Each row reports the share of a party bloc's legislator-terms (1910--1950) carrying an enduring-elite surname and the share carrying a common surname, with the number of legislator-terms in parentheses; the residual is the \emph{other} group. Right-of-center parties (Conservador, Liberal) are elite-heavy; left-of-center parties (Radical, Socialista, Comunista) are common-heavy. The left parties contain few, not zero, elite surnames. Source: 298-surname coding applied to the legislator roster (\texttt{cong\_df.csv}; replication repository, scripts \texttt{02}/\texttt{02b}).
\end{tablenotes}
\end{threeparttable}
\end{table}

\section{Robustness}\label{app-sec:robust}

The substantive conclusions are stable across measure variants, population references, and coding and modeling choices. This section reports the formal checks behind that claim: the robustness of the 1925 break, a survivorship lower bound on the early elite share, the sensitivity of the composition results to the socioeconomic cutoff and to a stricter historical (persistence-depth) gate, the sensitivity of the break to the business-leader route, the within-group agenda trajectories that show the contrast is inconsistent with a secular-shift confound, and a theme-level reading of the topic contrast that confronts the multiple-testing concern, and then summarizes the remaining checks.

\subsection{Robustness of the 1925 break}\label{app-subsec:itsrobust}

\begin{figure}[htbp]
\centering
\includegraphics[width=0.9\linewidth]{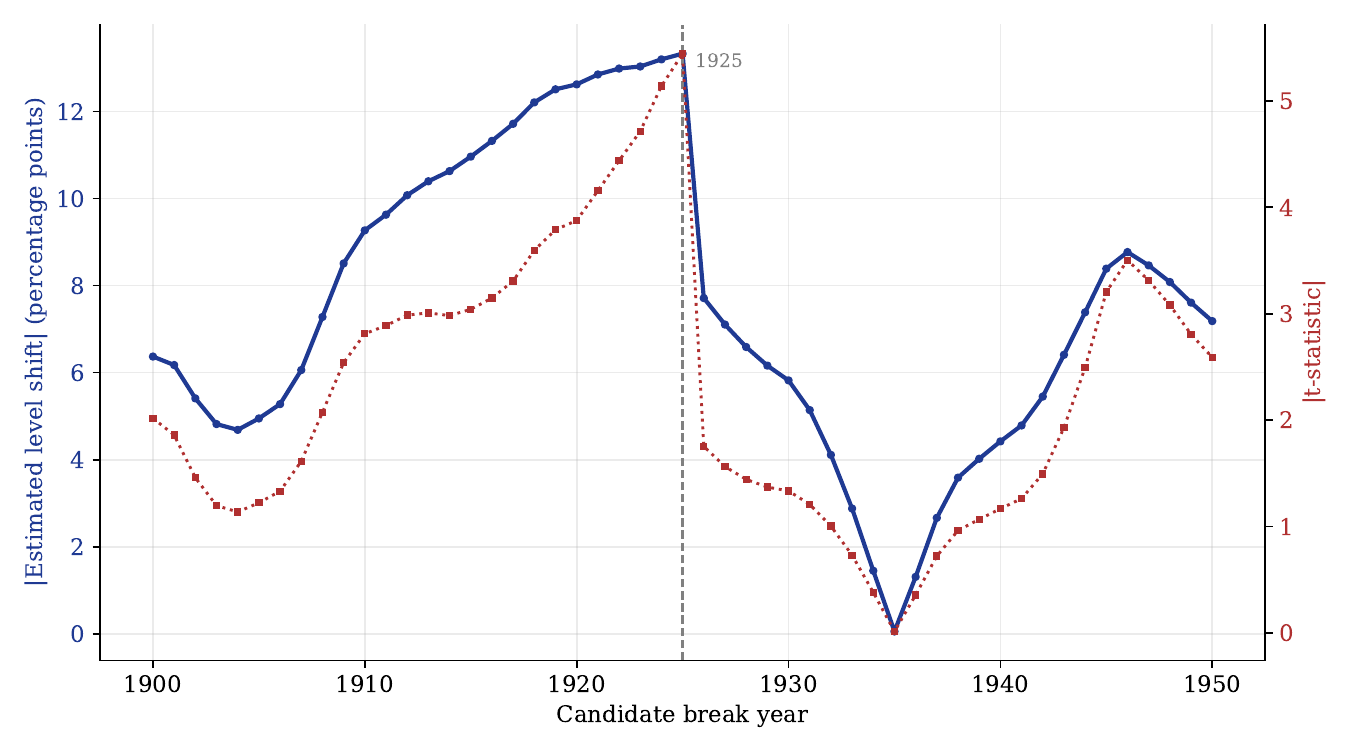}
\caption{Break-year placebo for the 1925 level shift. The figure plots the estimated level shift (in percentage points) from the interrupted-time-series regression as the break year is moved across each candidate year from 1900 to 1950, holding the estimation window fixed. Read the height of each bar as the magnitude of the discontinuity the model would attribute to a break at that year: the shift at 1925 is the largest in absolute value across all candidate years, so the 1925 step is the global maximum rather than one of several comparable breaks. Source: elite-surname coding of the full legislative roster; replication output for the break-year placebo.}
\label{app-fig:itsplacebo}
\end{figure}

The level shift at 1925 reported in the main text is not an artifact of the break year we chose or of the estimation window. \Cref{app-tab:itsrobust} sweeps the window: the level shift is $-13.80$ points for 1900--1955, $-13.45$ for 1890--1960, and $-13.32$ for 1880--1970, every estimate significant at $p<0.001$. The $-13.3$ step is stable to within half a point across windows that differ by decades. \Cref{app-fig:itsplacebo} runs the complementary placebo: it re-estimates the level shift with the break placed at each candidate year from 1900 to 1950. The shift at 1925 is the global maximum in absolute value across all candidate years, and it is the local maximum within every window. The break is not one of many comparable discontinuities the segmented model could have found; it is the single largest, and it sits at 1925.

\begin{table}[htbp]
\centering
\begin{threeparttable}
\caption{The 1925 level shift across estimation windows}\label{app-tab:itsrobust}
\small
\begin{tabular}{lccc}
\toprule
 & 1925 level shift & Std.\ error & $p$-value \\
Estimation window & (pts) & (HAC, 5 lags) & \\
\midrule
1900-1955 & -13.80*** & (2.23) & 0.000 \\
1890-1960 & -13.45*** & (2.02) & 0.000 \\
1880-1970 & -13.32*** & (2.45) & 0.000 \\
\bottomrule
\end{tabular}

\begin{tablenotes}\small
\item \textit{Notes:} Each row re-estimates the interrupted-time-series regression of the elite surname share of Congress on a pre-break trend, a level shift at 1925, and a post-break trend, over the estimation window named in the first column. The reported coefficient is the 1925 level shift in percentage points. Standard errors are heteroskedasticity- and autocorrelation-consistent (HAC, five lags). The step is near $-13$ points and significant at $p<0.001$ in every window. *** $p<0.01$. Source: elite-surname coding of the full legislative roster; replication output for the interrupted-time-series robustness sweep.
\end{tablenotes}
\end{threeparttable}
\end{table}

\subsection{A survivorship lower bound on the early elite share}\label{app-subsec:survivorship}

\begin{figure}[htbp]
\centering
\includegraphics[width=0.9\linewidth]{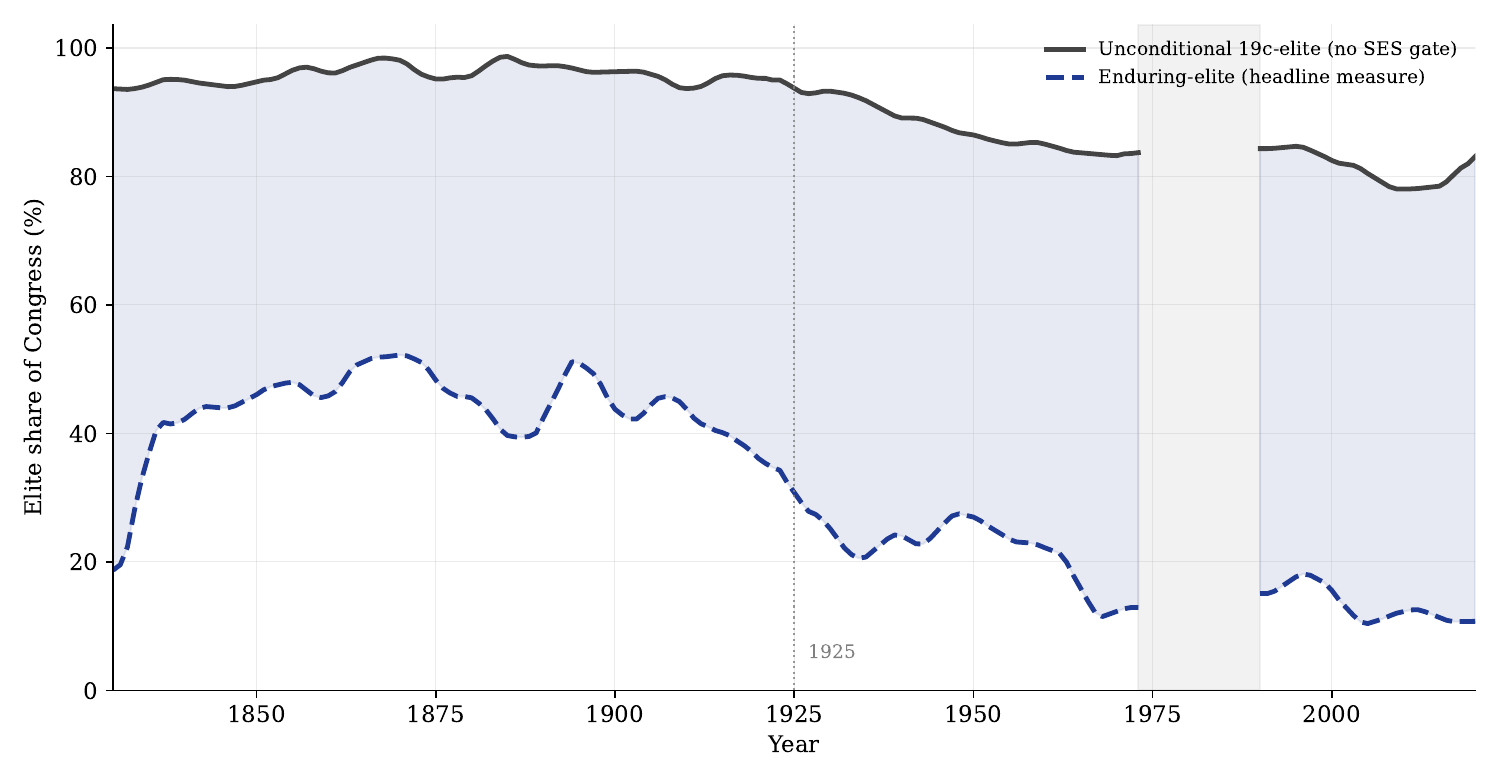}
\caption{Survivorship lower bound on the early elite share of Congress. The figure overlays the enduring-elite series (both criteria applied) on an unconditional nineteenth-century elite series built from the 1874 \emph{censo} top quartile and the nineteenth-century registers with no 2020 socioeconomic criterion (about $1{,}126$ surnames). Read the vertical gap between the two series as the amount the contemporary criterion removes: the unconditional series lies above the enduring series throughout the pre-1920 period (for example, $98\%$ versus $51\%$ in 1865 and $96\%$ versus $44\%$ in 1905) and never below it, so the enduring-elite trend is a strict lower bound on early elite presence. The unconditional list is permissive (it flags more than $93\%$ of early legislators), so the bound is real but loose. Source: enduring-elite and unconditional surname codings of the full legislative roster; replication output for the survivorship bound.}
\label{app-fig:survivorship}
\end{figure}

The enduring-elite measure requires a surname to clear a contemporary criterion as well as the nineteenth-century one. A natural worry is that this 2020 criterion discards nineteenth-century families that were elite then but have since faded, so that the early elite share of Congress is understated. \Cref{app-fig:survivorship} addresses the worry directly. It compares the enduring-elite series with an \emph{unconditional} nineteenth-century elite series, built from the 1874 \emph{censo} top quartile and the nineteenth-century registers with no 2020 socioeconomic criterion applied (about $1{,}126$ surnames). The unconditional series sits far above the enduring-elite series throughout the pre-1920 period ($98\%$ against $51\%$ in 1865, $96\%$ against $44\%$ in 1905), and in no year does it fall below the enduring series. Survivorship therefore attenuates the level of the enduring-elite series; it never inflates it. The enduring-elite trend is a strict lower bound on early elite presence in Congress.

Two caveats keep this honest. First, the unconditional list is very permissive: it flags more than $93\%$ of legislators in the early period, so the bound it provides is real but loose: it shows the direction of the bias, not a tight ceiling. Second, the bound is established by direct comparison of the two series rather than by a clean subset guarantee, because the unconditional list is rebuilt deterministically from the historical sources (the reconstruction recovers $287$ of the $298$ published surnames). The lower-bound claim is qualitative: the 2020 criterion can only have lowered the early elite share, never raised it. It is not a precise quantification of how much was lost.

\subsection{Sensitivity to the socioeconomic cutoff}\label{app-subsec:cutoff}

The contemporary criterion admits a surname whose socioeconomic index is at least $65$. \Cref{app-tab:cutoff} shows that the headline composition results do not turn on that threshold. Lowering the cutoff to $60$ or raising it to $70$ changes the size of the surname list (from $446$ to $241$ surnames) and, mechanically, the level of the elite share, but it leaves the pattern intact: the 1860s elite share runs $58.5$, $49.1$, and $33.2$ percent across the three cutoffs; the 2010s share runs $15.8$, $11.1$, and $9.8$ percent; the over-representation around 1900 runs $23.4$, $19.5$, and $14.2$-fold; and the 1925 level shift runs $-15.1$, $-11.3$, and $-10.5$ points, all significant at $p<0.001$. The published $298$-surname measure reproduces the main-text headline exactly. The sensitivity rows in \cref{app-tab:cutoff} use a deterministic reconstruction of the historical criterion, while the published row uses the deposited surname list; the two agree closely.

\begin{table}[htbp]
\centering
\begin{threeparttable}
\caption{Sensitivity of the composition results to the socioeconomic cutoff}\label{app-tab:cutoff}
\small
\begin{tabular}{lccccc}
\toprule
 & Surname & Elite share & Elite share & Over-rep. & 1925 level \\
Specification & count & 1860s (\%) & 2010s (\%) & fold $\sim$1900 & break (pts) \\
\midrule
Published measure (SES $\geq$ 65) & 298 & 51.2 & 11.3 & 21.1$\times$ & -13.32*** \\
\\[0.3em]
\multicolumn{6}{l}{\textit{Panel A: SES cutoff, gate-2 reconstructed (bearer floor = 30)}} \\
SES $\geq$ 60 & 446 & 58.5 & 15.8 & 23.4$\times$ & -15.13*** \\
SES $\geq$ 65 & 345 & 49.1 & 11.1 & 19.5$\times$ & -11.29*** \\
SES $\geq$ 70 & 241 & 33.2 & 9.8 & 14.2$\times$ & -10.51*** \\
\\[0.3em]
\multicolumn{6}{l}{\textit{Panel B: bearer floor (SES $\geq$ 65)}} \\
$\geq$ 5 bearers & 433 & 51.6 & 11.1 & 21.0$\times$ & -11.97*** \\
$\geq$ 30 bearers & 345 & 49.1 & 11.1 & 19.5$\times$ & -11.29*** \\
$\geq$ 50 bearers & 304 & 47.6 & 11.1 & 19.0$\times$ & -11.38*** \\
\bottomrule
\end{tabular}

\begin{tablenotes}\small
\item \textit{Notes:} Each row applies the enduring-elite construction with a different contemporary criterion and reports the resulting surname count, the elite share of Congress in the 1860s and the 2010s, the over-representation of the elite among legislators relative to the population around 1900, and the 1925 level shift from the interrupted-time-series regression. Panel A varies the socioeconomic-index cutoff (60/65/70); Panel B varies the minimum bearer count. The first row is the published $298$-surname measure (deposited list); the remaining rows use a deterministic reconstruction of the historical criterion. The headline pattern (high early share, sharp decline, large over-representation, and a significant 1925 break) holds across every variant. *** $p<0.01$. Source: enduring-elite codings under alternative cutoffs; replication output for the cutoff-sensitivity sweep.
\end{tablenotes}
\end{threeparttable}
\end{table}

\subsection{A stricter historical gate (persistence depth)}\label{app-subsec:depthgate}

The historical criterion admits a surname that appears in \emph{at least one} nineteenth-century register. A natural worry is that this disjunctive gate is too permissive, letting through surnames attested in only a single source. The persistence-depth score (\cref{app-sec:construction}) counts the registers in which a surname appears, so it supports a stricter re-run. \Cref{app-tab:depthgate} restricts the measure to the $102$ surnames attested in \emph{two or more} registers (depth $\geq 2$) and re-estimates the composition trend and the 1925 break. The break holds: the elite share falls from $42.7\%$ in the 1860s to $7.6\%$ in the 2010s, and the 1925 level shift is $-13.76$ points (HAC $p<0.001$). We read this narrowly. A larger pre-to-post step under the stricter gate is close to mechanical rather than independent corroboration, because the depth-$\geq 2$ gate re-selects exactly the families most concentrated in Congress before 1925 and most depleted after, so a steeper drop is partly what the selection builds in. The claim the gate licenses is the narrow one: the 1925 break is robust to dropping singly-attested surnames. The monotonicity panel of \cref{app-tab:depthmono} supports it. As the gate tightens from depth $\geq 1$ to $\geq 4$, the level shift stays negative and highly significant at every step ($-13.32$, $-13.76$, $-9.49$, $-7.44$ points, all HAC $p<0.001$), attenuating only as the surname base shrinks from $298$ to $37$ names (\texttt{scripts/33\_depth\_monotonicity.py}, output \texttt{m4\_depth\_monotonicity.csv}). The depth $\geq 1$ row is the full canonical $298$-surname list, whose break is the headline $-13.32$ points; this base, and the attenuating series above it, should not be confused with the $-11.13$-point break obtained when the thirteen business-route surnames are dropped from that list (\cref{app-subsec:bizroutrobust}). The break is therefore not an artifact of singly-attested surnames, though the deepest gate is not, on its own, stronger evidence than the baseline.

\begin{table}[htbp]
\centering
\begin{threeparttable}
\caption{The composition trend and the 1925 break under a stricter historical gate}\label{app-tab:depthgate}
\small
\begin{tabular}{lcccc}
\toprule
Gate & Surnames & 1860s share & 2010s share & 1925 level shift \\
\midrule
Both criteria (depth $\geq$ 1) & 298 & 49.9 & 11.6 & -13.3 \\
Stricter historical gate (depth $\geq$ 2) & 102 & 42.7 & 7.6 & -13.8 \\
\bottomrule
\end{tabular}

\begin{tablenotes}\small
\item \textit{Notes:} Each row applies the enduring-elite construction with a different historical gate and reports the elite share of Congress in the 1860s and the 2010s and the 1925 level shift from the interrupted-time-series regression (HAC standard errors, five lags). The first row is the published measure (a surname clears the historical criterion if it appears in at least one nineteenth-century register, depth $\geq 1$, $298$ surnames); the second restricts to surnames appearing in two or more registers (depth $\geq 2$, $102$ surnames). The long decline and the significant 1925 break hold, slightly strengthened, under the stricter gate. *** $p<0.001$. Source: enduring-elite coding under alternative historical gates; replication output \texttt{replication\_repo/output/m4\_depth\_gate.csv}.
\end{tablenotes}
\end{threeparttable}
\end{table}

\begin{table}[htbp]
\centering
\begin{threeparttable}
\caption{The 1925 break across persistence-depth gates (monotonicity)}\label{app-tab:depthmono}
\small
\begin{tabular}{lcccc}
\toprule
Gate & Surnames & 1860s share & 2010s share & 1925 level shift \\
\midrule
Depth $\geq 1$ (published) & 298 & 49.9 & 11.6 & $-13.32^{***}$ \\
Depth $\geq 2$             & 102 & 42.7 &  7.6 & $-13.76^{***}$ \\
Depth $\geq 3$             &  52 & 33.9 &  5.1 & $-9.49^{***}$ \\
Depth $\geq 4$             &  37 & 25.4 &  5.0 & $-7.44^{***}$ \\
\bottomrule
\end{tabular}
\begin{tablenotes}\small
\item \textit{Notes:} Each row re-runs the enduring-elite construction keeping only surnames attested in at least the stated number of nineteenth-century registers, and reports the elite share of Congress in the 1860s and the 2010s and the 1925 level shift from the interrupted-time-series regression (HAC standard errors, five lags). The base is the full canonical $298$-surname list (depth $\geq 1$ row, break $-13.3$), not the $285$-surname list obtained by dropping the business-route surnames (break $-11.1$; \cref{app-subsec:bizroutrobust}). The level shift stays negative and significant at every gate, attenuating as the surname base shrinks, so the break is not a single favorable slice. The deeper gates re-select the families most concentrated in Congress before 1925, so a steeper drop there is partly mechanical; the panel's role is to show the break survives dropping singly-attested surnames, not to corroborate the magnitude. *** $p<0.001$. Source: replication script \texttt{scripts/33\_depth\_monotonicity.py}; output \texttt{m4\_depth\_monotonicity.csv}.
\end{tablenotes}
\end{threeparttable}
\end{table}

\subsection{Sensitivity to the surname coding rule}\label{app-subsec:codingrule}

A legislator carries two surnames, so the categorical elite/common coding requires a rule for combining them. The main text uses the permissive rule: a legislator is coded elite if \emph{either} surname is on the enduring-elite list. \Cref{app-tab:codingrule} shows that the composition trend and the 1925 break survive two alternatives. Under the permissive rule the 1925 level shift is $-13.32$ points ($p<0.001$); under a strict rule that requires \emph{both} surnames to be elite it is $-2.85$ points ($p=0.004$); and under a paternal-only rule it is $-7.58$ points ($p<0.001$). Under all three codings the break is negative and significant, and the long decline is present throughout: the elite share in 1865, 1905, and 2010 runs $52.0$, $43.6$, and $12.3$ percent under the permissive rule, $14.0$, $12.9$, and $2.9$ percent under the strict rule, and $32.0$, $27.1$, and $8.7$ percent under the paternal-only rule. The strict break is mechanically smaller because the strict elite base rate is far lower (about $14\%$ versus about $52\%$ in 1865), not because the break weakens: its sign and significance are preserved. The choice of rule moves the level of the series but not the qualitative result.

\begin{table}[htbp]
\centering
\begin{threeparttable}
\caption{The composition trend and the 1925 break under alternative surname coding rules}\label{app-tab:codingrule}
\small
\begin{tabular}{lccccc}
\toprule
 & 1865 & 1905 & 2010 & 1925 break & $p$-value \\
Coding rule & share (\%) & share (\%) & share (\%) & (pt) & \\
\midrule
Permissive (OR) & 52.0 & 43.6 & 12.3 & -13.32 & $<$0.001 \\
Strict (AND) & 14.0 & 12.9 & 2.9 & -2.85 & 0.004 \\
Paternal only & 32.0 & 27.1 & 8.7 & -7.58 & $<$0.001 \\
\bottomrule
\end{tabular}

\begin{tablenotes}\small
\item \textit{Notes:} Each row applies a different rule for combining a legislator's two surnames into a categorical elite coding and reports the resulting elite share of Congress in 1865, 1905, and 2010, together with the 1925 level shift from the interrupted-time-series regression and its $p$-value. The permissive (OR) rule codes a legislator elite if either surname is on the enduring-elite list (the main-text rule); the strict (AND) rule requires both surnames to be elite; the paternal-only rule uses the paternal surname alone. The break is negative and significant under all three rules; the smaller strict break reflects the much lower strict elite base rate, not a weaker break. Source: enduring-elite coding of the full legislative roster under alternative combination rules; replication output \texttt{replication\_repo/output/coding\_rule\_sensitivity.csv}.
\end{tablenotes}
\end{threeparttable}
\end{table}

\subsection{Sensitivity to the business-leader route}\label{app-subsec:bizroutrobust}

Thirteen surnames enter the measure through the business-leader route (Route 1b) rather than the socioeconomic index (\cref{app-subsec:bizroute}). Because that route applies two thresholds (an over-representation ratio of at least eight and at least five appearances among business leaders), a natural question is whether the headline 1925 break depends on them or on those particular cutoffs. It does not. Dropping all thirteen business-route surnames, so that the elite list reduces to the padr\'on-plus-historical surnames alone, moves the 1925 level shift from $-13.32$ to $-11.13$ points: a movement of about two points, still large and significant (HAC $p<0.001$). The 1860s elite share moves only from $49.9$ to $47.1$ percent and the 2010s share is essentially unchanged at about $11.6$ percent, so the long decline is untouched. Sweeping the two thresholds jointly (over-representation in $\{6,8,12\}$ and appearances in $\{3,5,10\}$, rebuilding the elite list deterministically in each of the nine cells and re-estimating the break) keeps the level shift in a narrow band from about $-12.3$ to $-13.3$ points, significant at $p<0.001$ in every cell. The two heaviest dynasties the route admits, Portales and Balmaceda, carry many legislator-terms (twenty-eight and fifty-six, of which twenty-five and forty-eight fall before 1925), yet the break is insensitive to whether they are in or out: starting from the full $298$-surname list (break $-13.32$), dropping only Portales and Balmaceda moves the 1925 level shift to $-12.24$ points (HAC $p<0.001$), and dropping those two plus Bulnes and Garc\'ia-Huidobro gives $-11.64$ points ($p<0.001$), so the $-13.3$ magnitude is not a two-family artifact (\texttt{scripts/32\_two\_dynasty\_leaveout.py}, output \texttt{m2\_two\_dynasty\_leaveout.csv}). The 1925 step is a property of the composition trend, not of the business-leader route or its thresholds (\texttt{scripts/29\_business\_route\_robustness.py}).

\subsection{Theme-level topic contrast and multiplicity}\label{app-subsec:themes}

The structural topic model produces $80$ topics. Two inferential concerns attend the topic-level estimates: the non-independence of a legislator's speeches, which inflates document-level $p$-values (\cref{app-sec:stm}), and multiplicity (with $80$ topics tested, some will cross any single-topic threshold by chance). The theme-level contrast reported as Table~1 in the main text confronts both: it aggregates the $80$ topics into coherent themes, which addresses the multiplicity, and estimates each theme effect with standard errors clustered by legislator, which corrects the non-independence. The elite and common emphasis on each theme follow. The elite-leaning themes are elections and parties ($+1.13$), defense ($+0.43$), state administration ($+0.38$), and foreign relations ($+0.26$); the common-leaning themes are labor and industry ($-1.77$), public works ($-0.56$), justice ($-0.41$), social welfare ($-0.31$), and local government ($-0.30$). Finance and taxation, often presumed an elite preoccupation, is mildly common-leaning ($-0.29$) at the theme level. The thematic clusters are the statecraft (elections, defense, administration, foreign relations) and labor--local-infrastructure divisions reported in the main text.

The stars in Table~1 sit on the emphasis-versus-\emph{other} columns, not on the elite-minus-common contrast the paper interprets. Computing the legislator-clustered significance on the contrast column itself (\texttt{scripts/35\_theme\_contrast\_pvalues.R}, output \texttt{m1\_contrast\_pvalues.csv}), only three of the eleven theme contrasts clear the $5\%$ threshold: labor and industry ($p=4\times10^{-7}$), foreign relations ($p=0.001$), and elections and parties ($p=0.026$); the remaining eight, including defense, state administration, public works, local government, social welfare, justice, finance, and education, do not, and two of these three no longer clear $5\%$ once the topic model's own estimation uncertainty is propagated (\cref{app-subsec:stminference}), so that no single theme contrast is load-bearing. This is why the object of interest is the two-cluster ranking of Table~1, not any single theme's star, and why the agenda claim is carried inferentially by the within-group slope tests of \cref{app-subsec:withinelite}, which survive clustering and topic-model uncertainty, rather than by the cross-sectional theme contrasts.

The multiplicity is stated plainly rather than hidden. Of the $80$ topics, $33$ reach $p<0.05$ on the elite side and $45$ on the common side at the (inflated) document level, against roughly four expected by chance under the null. No single topic's $p$-value carries the claim. What carries it is the thematic coherence of the two clusters: the topics that separate the groups are not a scatter of unrelated subjects but two recognizable agendas, and that coherence, not any one topic, is the evidence. The full theme-level table is Table~1 in the main text.

The assignment of topics to themes is reported in full in \cref{app-tab:themecrosswalk}, so the aggregation can be audited rather than taken on trust. Each of the $80$ topics is placed in exactly one theme by reading its leading and FREX terms (the labeling procedure of \cref{app-subsec:stmtopics}); $21$ topics that carry no interpretable policy content (procedural, rhetorical, or low-coherence OCR artifacts) fall in a residual \emph{Other} bucket that enters no theme contrast. The crosswalk is deposited as \texttt{topic\_theme\_map.csv} in the replication repository.

\begin{table}[htbp]
\centering
\begin{threeparttable}
\caption{Assignment of structural-topic-model topics to substantive themes}\label{app-tab:themecrosswalk}
\small
\begin{tabular}{lcL{8.2cm}}
\toprule
Theme & \# topics & Constituent topic numbers \\
\midrule
Elections \& parties & 3 & 7, 58, 73 \\
Defense \& security & 5 & 23, 29, 42, 61, 75 \\
State administration & 5 & 1, 20, 24, 37, 68 \\
Foreign relations & 1 & 45 \\
Education & 1 & 65 \\
Finance \& taxation & 12 & 4, 5, 6, 12, 15, 27, 32, 33, 36, 48, 53, 77 \\
Social welfare \& health & 6 & 8, 28, 50, 62, 70, 74 \\
Local government \& land & 6 & 10, 11, 30, 43, 55, 56 \\
Justice \& law & 5 & 25, 39, 40, 46, 71 \\
Public works & 6 & 2, 19, 22, 35, 47, 60 \\
Labor \& industry & 9 & 16, 41, 44, 49, 51, 59, 63, 66, 78 \\
Other & 21 & 3, 9, 13, 14, 17, 18, 21, 26, 31, 34, 38, 52, 54, 57, 64, 67, 69, 72, 76, 79, 80 \\
\bottomrule
\end{tabular}

\begin{tablenotes}\small
\item \textit{Notes:} Each of the $80$ topics from the structural topic model \parencite{Roberts2014} is assigned to exactly one theme by reading its highest-probability and FREX terms (\cref{app-subsec:stmtopics}). Topic numbers are the model's own indices and match those in \cref{app-tab:stmcoefs}. The eleven substantive themes are the units of the main-text contrast (Table~1); the residual \emph{Other} bucket collects procedural, rhetorical, and low-coherence OCR topics that carry no agenda content and enter no theme contrast. Source: replication input \texttt{topic\_theme\_map.csv}.
\end{tablenotes}
\end{threeparttable}
\end{table}

\subsection{Within-group agenda trajectory, 1910--1949}\label{app-subsec:withinelite}

The main-text agenda contrast could in principle be an artifact of the period's broader mobilization rather than of legislative composition. The 1910--1950 window brought the FOCH labor federation, the Socialist Workers' Party of Recabarren, and the Alessandri election of 1920, any of which could have pulled the whole chamber toward labor and social questions. If that secular opening, and not who sat in Congress, drove the contrast, the labor-and-industry emphasis should rise in parallel for elite- and common-surname legislators alike, leaving the gap flat.

It does not. \cref{app-tab:withinelite} traces the labor-and-industry theme emphasis decade by decade within each surname group. It rises for both, consistent with a genuine broadening of the agenda, but it rises nearly twice as fast among common-surname legislators, so the elite--common gap widens from about $0.3$ points in the 1910s to about $3.6$ points in the 1940s. Fitting the theme prevalence on a linear decade trend within each group, with party-bloc and chamber controls and standard errors clustered by legislator, the adjusted slope is $+0.80$ points per decade for elite legislators ($\mathrm{SE}=0.22$) and $+1.53$ for common legislators ($\mathrm{SE}=0.20$). A pooled interaction puts the difference in slopes at $+0.90$ points per decade ($\mathrm{SE}=0.29$, $p=0.002$), net of party and chamber. The compositional contrast is therefore not a by-product of the secular agenda shift: as common surnames take seats, the floor turns toward labor faster than the period's general drift, and the gap between the two groups grows rather than closes (\texttt{scripts/27\_within\_elite\_trajectory.R}).

This slope difference is the one piece of the agenda analysis that does inferential work against a confound, so we confirm it survives the topic model's own estimation uncertainty. Re-estimating it by the method of composition, drawing fifty samples of the topic proportions from their posterior and combining them by Rubin's rules (\cref{app-subsec:stminference}, \texttt{scripts/28\_stm\_uncertainty\_methodofcomposition.R}), holds the common-minus-elite slope difference at $+0.84$ points per decade (combined $\mathrm{SE}=0.29$, $p=0.004$); the topic-estimation component is only about $7\%$ of the total variance, so the result does not rest on treating estimated topic proportions as observed.\footnote{Re-running the method of composition at a second seed and at $B=200$ draws holds the common-minus-elite slope difference at $+0.83$ to $+0.84$ points per decade (combined $\mathrm{SE}\approx0.29$, $p\approx0.004$), with the topic-estimation variance share about $5$ to $7\%$, confirming the result is insensitive to seed and draw count (\texttt{scripts/36\_seed\_robustness.R}, output \texttt{should3\_seed\_robustness.csv}).} This propagation, like the one in \cref{app-subsec:stminference}, carries the document--topic ($\theta$) uncertainty conditional on the fitted model, not uncertainty in $\beta$, the topic count, or the initialization; the theme aggregation and $K$-insensitivity argument cover those choices separately.

\begin{table}[htbp]
\centering
\begin{threeparttable}
\caption{Labor-and-industry emphasis by surname group and decade, 1910--1949}\label{app-tab:withinelite}
\small
\begin{tabular}{lccc}
\toprule
Decade & Elite & Common & Gap (Common $-$ Elite) \\
\midrule
1910s & 6.6 & 6.9 & 0.3 \\
1920s & 7.5 & 10.1 & 2.6 \\
1930s & 9.8 & 11.3 & 1.6 \\
1940s & 8.7 & 12.3 & 3.6 \\
\bottomrule
\end{tabular}

\begin{tablenotes}\small
\item \textit{Notes:} Cell entries are the mean structural-topic-model prevalence of the labor-and-industry theme, in percentage points, across floor speeches by legislators in each surname group and decade (\cref{app-subsec:themes}). The gap column is the common-minus-elite difference. Within-group party- and chamber-adjusted decade slopes are $+0.80$ (elite) and $+1.53$ (common) points per decade; the difference in slopes is $+0.90$ ($p=0.002$), with standard errors clustered by legislator. The rising common slope and widening gap show that the main-text agenda contrast is not an artifact of the period's general agenda opening. Source: replication script \texttt{scripts/27\_within\_elite\_trajectory.R}; outputs \texttt{within\_elite\_trajectory.csv} and \texttt{within\_elite\_trajectory\_means.csv}.
\end{tablenotes}
\end{threeparttable}
\end{table}

A mirror-image test confirms the reading. If the agenda contrast tracks composition rather than the period's general drift, the \emph{elite} advantage on the statecraft themes (elections and parties, defense, state administration, foreign relations) should \emph{widen} as elites concentrate in a shrinking set of seats, not narrow. It widens. Fitting the same within-group decade trend to the statecraft bundle, elite-surname legislators accelerate toward statecraft at $+1.35$ points per decade ($\mathrm{SE}=0.25$), while common-surname legislators are flat ($+0.19$, $\mathrm{SE}=0.32$, not significant); the difference in slopes is $+1.34$ points per decade ($\mathrm{SE}=0.40$, $p<0.001$), net of party and chamber (\cref{app-tab:statecraft}). The two agendas therefore move apart symmetrically across the window: common surnames pull the floor toward labor while elite surnames pull it toward statecraft. The widening shows at both poles of the contrast, as a compositional reading predicts and a single secular shift would not (\texttt{scripts/31\_statecraft\_trajectory.R}).

\begin{table}[htbp]
\centering
\begin{threeparttable}
\caption{Statecraft-theme emphasis by surname group and decade, 1910--1949}\label{app-tab:statecraft}
\small
\begin{tabular}{lccc}
\toprule
Decade & Elite & Common & Gap (Elite $-$ Common) \\
\midrule
1910s & 13.1 & 13.0 & 0.1 \\
1920s & 15.7 & 14.6 & 1.1 \\
1930s & 13.7 & 13.6 & 0.1 \\
1940s & 17.7 & 15.1 & 2.5 \\
\bottomrule
\end{tabular}

\begin{tablenotes}\small
\item \textit{Notes:} Cell entries are the mean structural-topic-model prevalence of the statecraft bundle (elections and parties, defense and security, state administration, and foreign relations), in percentage points, across floor speeches by legislators in each surname group and decade. The gap column is the elite-minus-common difference. Within-group party- and chamber-adjusted decade slopes are $+1.35$ (elite) and $+0.19$ (common, not significant) points per decade; the difference in slopes is $+1.34$ ($p<0.001$), with standard errors clustered by legislator. The widening elite advantage on statecraft mirrors the widening common advantage on labor and industry (\cref{app-tab:withinelite}). Source: replication script \texttt{scripts/31\_statecraft\_trajectory.R}; outputs \texttt{statecraft\_trajectory.csv} and \texttt{statecraft\_trajectory\_means.csv}.
\end{tablenotes}
\end{threeparttable}
\end{table}

\subsection{Further checks}

The remaining checks are summarized here.

\begin{itemize}[topsep=2pt,itemsep=3pt]
  \item \textbf{Measure variants.} The discourse and composition results are qualitatively stable across three versions of the measure: the original $65$-surname land-only list (1874 census alone), the $285$-surname padr\'on-plus-historical list (a superseded intermediate), and the canonical $298$-surname list that adds the business-leader route. The group gaps shrink mechanically as the list broadens (a wider elite list dilutes the average elite signal), but the direction of every contrast holds.

  \item \textbf{Population reference.} The over-representation conclusion is unchanged whether the elite share of Congress is measured against a dynamic, person-level population reference (the civil registry, about $22.65$ million births, with the reference distribution allowed to shift over time) or against a static benchmark. If anything, the dynamic reference understates early over-representation less than the static one, so the headline over-representation is not an artifact of the reference choice.

  \item \textbf{Coding rules.} Both the paternal and the maternal surname are used to classify a legislator. When a legislator carries both an elite and a common surname, elite classification takes precedence over common. The results are not sensitive to this precedence rule: the composition trend and the 1925 break survive a strict (both-surnames) rule and a paternal-only rule (\cref{app-tab:codingrule}, \cref{app-subsec:codingrule}).

  \item \textbf{Structural topic model.} The topic contrast holds under both the party-, chamber-, and year-adjusted prevalence specification and the unconditional one. In both, the defense and finance topics fall on the elite side and the labor and local-infrastructure topics on the common side.

  \item \textbf{Text-based ideal points.} The continuous surname-status result is robust across specifications and to clustering by legislator or by family. The categorical common-surname effect is positive throughout but specification-sensitive in significance; the main text states this honestly rather than over-reading the categorical contrast.
\end{itemize}

\clearpage
\small

\clearpage
\small \printbibliography

\end{document}